\documentclass[10pt,fleqn]{lf-elsart}

\newcommand{\seq}[1]{\stackrel{#1}{\sim}}
\newcommand{\seqs}{\seq{\bfx}}

\newcommand{\seqst}{\seq{\bfx, t}}
\newcommand{\seqss}{\seq{\bfx,\bfxp}}
\newcommand{\seqsst}{\seq{\bfx, \bfxp, t}}

\usepackage{amssymb}
\usepackage{amsmath}
\newenvironment{proof}{\noindent\textbf{Proof:}}{\mbox{}\hfill\qed}
\RequirePackage{lfmath} \RequirePackage{citesort}

\begin{document}
\small
\begin{frontmatter}



\title{On the elimination of the sweeping interactions from theories of hydrodynamic turbulence}


\author[lf]{Eleftherios Gkioulekas\corauthref{cor1}}
\ead{lf@mail.ucf.edu}
\corauth[cor1]{Corresponding Author}
\address[lf]{Mathematics,  University of Central Florida, Orlando, FL 32816-1364}

\begin{abstract}
In this paper, we revisit the claim that the Eulerian and quasi-Lagrangian same time correlation tensors are equal. This statement  allows us to transform the results of an MSR quasi-Lagrangian statistical theory of hydrodynamic turbulence back to the Eulerian representation. We define a hierarchy of homogeneity symmetries between incremental homogeneity  and global homogeneity. It is shown that both the elimination of the sweeping interactions and the derivation of the $4/5$-law require a homogeneity assumption stronger than  incremental homogeneity but weaker than global homogeneity.  The quasi-Lagrangian transformation, on the other hand, requires an even stronger homogeneity assumption which is many-time rather than one-time but still weaker than many-time global homogeneity. We argue that it is possible to relax this stronger assumption and still preserve the conclusions derived from theoretical work based on the quasi-Lagrangian transformation.  
\end{abstract}

\end{frontmatter}

  \newtheorem{definition}{Definition}
  \newtheorem{proposition}{Proposition}
  \newtheorem{theorem}{Theorem}
  \newtheorem{corollary}{Corollary}
  \newtheorem{remark}{Remark}
  \newtheorem{exercise}{Exercise}

\section{Introduction} 
 
A remarkable feature of hydrodynamic turbulence in three dimensions is that it exhibits universal self-similarity properties at small length scales independently of the forcing mechanism that operates at larger length scales. The self-similar nature of turbulence was noticed by Richardson \cite{book:Richardson:1922} who suggested that large vortices will generate increasingly smaller vortices until they become hydrodynamically stable and then get dissipated by viscosity.  Kolmogorov \cite{article:Kolmogorov:1941,article:Kolmogorov:1941:1} conjectured that for length scales $r$ between the forcing scale $\ell_0$ and the dissipation scale $\eta$, the structure functions $S_n (\bfx, r\bfe)$ will be independent of $\ell_0$ and $\eta$, and, as was pointed out by Batchelor \cite{article:Batchelor:1947}, this conjecture implies that $S_n (\bfx, r\bfe)$ satisfy the following power laws:
\begin{equation}
S_n (\bfx, r\bfe) = \avg{\{[\bfu (\bfx + r\bfe, t) - \bfu (\bfx, t)]\cdot\bfe\}^n}  = C_n (\gee r)^{n/3}.
\end{equation}
Here $\bfe$ is a unit vector, and $\gee$ equals the rate of energy injection into the fluid, the energy flux in the cascade of energy from large scales to small scales, and the rate of energy dissipation at small scales.  The constant $C_n$ was believed to be universal, but in fact it is not (except for $n=3$) and  it is dependent on the forcing spectrum.  From the above, the energy spectrum $E(k)$ for $\ell_0^{-1} \ll k \ll \gn^{-1}$ can be shown to satisfy 
\begin{equation}
E (k) = C \gee^{2/3} k^{-5/3}.
\end{equation}
This prediction was confirmed for the first time in 1962 \cite{article:A.Moilliet:1962,article:Gibson:1962}, and today, with modern computers, it is routinely reproduced in numerical simulations. It has since come to light \cite{book:Frisch:1995,article:Antonia:1997} that there exist departures from Kolmogorov scaling laws for the higher order structure functions (known as intermittency corrections), and Kolmogorov  (with Oboukhov) \cite{article:Kolmogorov:1962,article:Oboukhov:1962} was in fact the first to propose revisions of his original theory.  The correct expression for $S_n (r)$  has the form 
\begin{equation}
S_n (r) = C_n (\gee r)^{n/3} (r/\ell_0)^{\gz_n - n/3}, 
\end{equation}
where $\gz_n$ are scaling exponents to be determined. The challenge here has been to develop theoretical understanding that can account for this energy cascade with a logical argument that begins from the underlying governing Navier-Stokes equations.  It is not only a matter of calculating the scaling exponents $\gz_n$.  The robustness of the scaling of the energy spectrum needs to be explained, and the universality of the scaling exponents themselves is in fact still an open question.  

The energy cascade from large scales to small scales is driven by the nonlinear term of the Navier-Stokes equations, and it is often explained as an effect of the vortex stretching  and tilting caused by that term.  However, the same term is also responsible for a sweeping interaction whereby a vortex is swept altogether from one location to another with minimal distortion.  Implicit in the idea of an energy cascade  is the assumption that these sweeping interactions have a negligible effect on  the structure functions in the inertial range.  It has therefore been necessary to use theoretical schemes that ``eliminate'' sweeping \cite{lect:Procaccia:1997}.  The goal of this paper is to call attention to the fact that these schemes do not \emph{prove} that sweeping is negligible; they only introduce the assumption that it is so.  Recent doubts \cite{article:Frisch:2005} concerning the consistency of the local homogeneity framework are directly linked with this problem of rigorously eliminating the sweeping interactions, and also with the problem of formulating a reasonable definition of local homogeneity. We will also make a conjecture, and explore its plausibility, which, if shown to be true, would establish the assumption that the sweeping interactions are negligible in the inertial range on a firmer ground.

It should be noted that a strictly rigorous mathematical theory based exclusively on the Navier-Stokes equations is a very difficult task.  For this reason, it is necessary to tolerate unproven assumptions as hypotheses, as long as such assumptions can be reasonably supported by physical arguments, or by experiment.  It is within a specific framework of reasonable assumptions, which will be defined in a moment, that we claim that sweeping elimination procedures still do not prove that sweeping interactions are negligible.  

The argument of this paper, summarily, is the following.  First, we show that the elimination of the sweeping interactions as well as the derivation of the $4/5$-law requires a homogeneity assumption stronger than the assumption of  incremental homogeneity, as envisioned by Frisch  \cite{book:Frisch:1995}.  Second, we show that using the quasi-Lagrangian formulation of Belinicher and L'vov to eliminate the sweeping interactions requires an even stronger homogeneity assumption which involves many-time correlations instead of one-time correlations. We conclude with a discussion of the implications of this argument on the utility of the quasi-Lagrangian formulation. Specifically, we will show that despite this apparent shortcoming of the quasi-Lagrangian formulation, the theoretical work based on it can still be used as a foundation for a physically useful theory, along the lines of the Frisch framework, provided that certain considerations are taken into account. Furthermore,  incremental homogeneity \emph{is} in fact a consistent framework, provided that the sweeping interactions can be eliminated in a more rigorous manner. 

The paper is organized as follows.  In section 2, we review the theoretical developments that gave rise to the issue of the sweeping interactions, and discuss the assumptions underlying most efforts to understand the energy cascade from a theoretical point of view.  The $4/5$-law is discussed in section 3 and the quasi-Lagrangian formulation in section 4. The implications of our argument for the theories that use the quasi-Lagrangian formulation as a foundation are discussed in section 5, and the paper is concluded in section 6.  Appendix \ref{sec:qlgoveqs} reviews how the quasi-Lagrangian formulation eliminates the sweeping interactions.  Appendix \ref{sec:det} presents a more complete account of the calculation of a functional determinant originally given by  L'vov and  Procaccia \cite{article:Procaccia:1995:1}. In  appendix \ref{app:gaussiansweep} we evaluate the contribution of the sweeping interactions in closed form for the case of passive random gaussian delta-time-decorrelated  sweeping.

\section{Theoretical background}

 We begin with reviewing the development of the ideas that form the theoretical foundation of certain  recent attempts to understand the universal behavior of turbulence.  The problem of the sweeping interactions and its resolution is an essential part of this theoretical foundation.  Then we discuss the set of assumptions that are widely accepted on physical grounds. Because the argument of this paper requires simultaneous consideration of a wide range of interdependent topical areas, this overview will help by providing the reader the  broader context against which the argument and its implications on the theoretical foundations of turbulence will be discussed later in our paper. This overview represents strictly my personal philosophical point of view. A more comprehensive and unbiased review of theoretical three-dimensional turbulence is  already available in the literature \cite{article:Nelkin:1994,book:Frisch:1995,lect:Procaccia:1997,article:Sreenivasan:1999}.

\subsection{Theoretical approaches to turbulence}

The foundation on which recent successful theoretical work was accomplished on the problem of the direct energy cascade rests on the following essential ideas:  The first critical idea is the framework of globally homogeneous and isotropic turbulence  introduced by Taylor \cite{article:Taylor:1935,article:Taylor:1936,article:Howarth:1938,article:Robertson:1940} and popularized by Batchelor \cite{book:Batchelor:1953}. Within that framework there have been numerous attempts to model turbulence using closure models \cite{book:Leslie:1972}.  The second critical idea, due to Kraichnan, is his discovery that such models are not realizable because they predict negative values for the energy spectrum \cite{article:Kraichnan:1957}.  Kraichnan counterproposed a different closure model \cite{article:Kraichnan:1958,article:Kraichnan:1959}, the direct interaction approximation (DIA), with the unique feature that it makes use of response functions.  Disagreement with experimental predictions prompted Kraichnan to call attention to the problem of sweeping interactions \cite{article:Kraichnan:1964}, and to revise his earlier model.  The new model \cite{article:Kraichnan:1965}, the Lagrangian history direct interaction approximation (LHDIA), was one of the first models to make predictions  in agreement with experiment \cite{article:Kraichnan:1966}.  A review of Kraichnan's work was given by Leslie \cite{book:Leslie:1972}. It is fair to say that LHDIA was the first successful theory of three-dimensional turbulence. Unfortunately, it was not clear how to generalize LHDIA, which was a first order approximation, to higher orders, and as a result, further development of this theoretical program was not possible for many years. 

Parallel to these efforts, there have also been attempts to construct exact mathematical theories of turbulence based on functional calculus.  The first such formulation was given by Hopf \cite{article:Hopf:1952}, and an equivalent reformulation in terms of path integrals by Rosen \cite{article:Rosen:1960,article:Rosen:1960:1}. Novikov \cite{article:Novikov:1965} modified the Hopf formalism to include a gaussian delta  correlated stochastic forcing, intended to model the hydrodynamic instability responsible for turbulence.  An interesting application of this formalism is the more rigorous and powerful reformulations of the original dimensional analysis arguments used by Kolmogorov \cite{article:Yanovskii:1976,article:Sazontov:1979}. Its main disadvantage is that it restricts the statistical description to one-time velocity correlations.  A generalization to include many-time velocity correlations was given by Lewis and Kraichnan \cite{article:Kraichnan:1962:1};  however even that is inadequate because it does not include response functions.  

The essential idea of the definitive approach was introduced by  Wyld \cite{article:Wyld:1961}.  The main result is that Feynman diagrams can be used to generalize DIA to higher orders, and that DIA itself is essentially a one-loop line-renormalized diagrammatic theory.  A generalization of this scheme to a wider range of dynamical systems was given by Martin, Siggia, and Rose \cite{article:Rose:1973}, although, as they themselves explained, without a sufficiently rigorous justification.  Phythian \cite{article:Phythian:1977} used Feynman path integrals to reformulate the MSR theory, and showed that it can be justified for dynamical systems that are local in time and first-order in time.  An assumption implicit in this argument is that the dynamical system has a unique solution for all time.  This claim has not been proven for the Navier-Stokes equations in three dimensions, however global regularity, as a matter of fact, can be proved rigorously \cite{article:Pavlovic:2002,article:Li:2005} if  the diffusion term in the Navier-Stokes equations is replaced with a hyperdiffusion term like $\nu\del^4 u_\ga$.  A pedagogical introduction to MSR theory was given recently by L'vov and  Procaccia \cite{lect:Procaccia:1994} and Eyink \cite{article:Eyink:1996:2}, and a careful review of the mathematical foundations of the theory itself is given in the paper by Andersen \cite{article:Andersen:2000} (also see references therein). 

 Unfortunately, the MSR formalism could not be applied to generalize Kraichnan's more successful LHDIA theory because the Navier-Stokes equations in the Lagrangian representation are not local in time.  Eventually, a way was discovered around this difficulty, thus breaking the deadlock that has been plagueing theory for decades. It involves  combining the MSR formalism with  renormalization schemes that eliminate the sweeping interactions.  The first such scheme was introduced by Yakhot \cite{article:Yakhot:1981}, and another by Belinicher and L'vov \cite{article:Lvov:1987,article:Lvov:1991}.  Combined with the MSR formalism, one has a rather solid foundation for further theoretical work.  It is these schemes, and the nature of the assumptions that they implicitly introduce, that will concern us in this paper.  

Since 1995, there have been some very remarkable developments in this direction:  L'vov and  Procaccia have used the quasi-lagrangian renormalization scheme \cite{article:Lvov:1987,article:Lvov:1991} to formulate a diagrammatic theory \cite{article:Procaccia:1995:1,article:Procaccia:1995:2,article:Procaccia:1996} that generalized Kraichnan's DIA to all orders.  It was shown that as long as the theory is truncated to finite order, it predicts agreement with Kolmogorov's theory and the absence of intermittency corrections \cite{article:Procaccia:1995:1}.  It was also shown that if the theory is not truncated, there is a critical divergence that \emph{does} lead to intermittency corrections \cite{article:Procaccia:1995:2}.  L'vov and  Procaccia \emph{et al} also formulated a nonperturbative theory \cite{article:Procaccia:1996:1,article:Procaccia:1996:2,article:Procaccia:1996:3,article:Procaccia:1997} based on the fusion rules which are predicted by the underlying diagrammatic theory.  This theory has been used to derive a nonperturbative method \cite{article:Procaccia:1998,article:Procaccia:1998:1,article:Procaccia:1998:2} and a perturbative method   \cite{article:Procaccia:2000} for calculating the scaling exponents $\gz_n$. The  perturbative method  \cite{article:Procaccia:2000} has been used successfully to calculate $\gz_n$ for all $n$ accessible to experimental measurement, but it requires that the deviation of $\gz_2$ from the Kolmogorov prediction $2/3$, which is the small parameter, be already known.  This ability of the L'vov-Procaccia theory to predict the existence of intermittency is a significant accomplishment. A partial review of  these developments was given in \cite{lect:Procaccia:1997}. The non-perturbative theory has also led to a clearer understanding of local isotropy  \cite{article:Procaccia:1999,article:Procaccia:2005}.

It is worth mentioning that there exists an entirely different theoretical approach to the problem based on renormalization group methods.  A detailed review is given in \cite{book:McComb:1990,book:Frisch:1995,article:Woodruff:1998} and some relevant criticism in Refs. \cite{article:Kraichnan:1987,article:Eyink:1994}. There are two interesting points of convergence between renormalization group methods and the theories reviewed previously.  First, Eyink \cite{article:Eyink:1993,article:Eyink:1993:1} employed the renormalization group method to derive the fusion rules under certain assumptions both for shell models of turbulence and for hydrodynamic turbulence itself.  The fusion rules are a crucial element in both the perturbative and the non-perturbative theories of L'vov and  Procaccia.  Second, Giles \cite{article:Giles:2001} used the renormalization group method to calculate the scaling exponents $\gz_n$, without relying on any experimental input, contrary to the paper \cite{article:Procaccia:2000}.  In this calculation, the sweeping interactions were eliminated using the scheme by Yakhot \cite{article:Yakhot:1981}.  A comparative study of the two approaches would help  further progress. 

\subsection{The  hypotheses that underlie MSR theory}
\label{sec:hypothesis}

In all the theoretical work that has been reviewed above, it is assumed that the Navier-Stokes equations have a unique solution, that there exists hydrodynamic instability leading to turbulence, and that this instability can be modeled with stochastic forcing acting at large scales. These assumptions are introduced implicitly simply by employing the MSR formalism.  Although they are widely accepted on physical grounds, there has also been substantial effort to deal with  them rigorously.  

An overview of the mathematical results on the existence and uniqueness of solutions to the Navier-Stokes equations is given in ref.~\cite{book:Temam:1999,lect:Galdi:2000} and references therein.  Briefly, in two dimensions the existence and uniqueness of strong solutions has been shown rigorously.  In three dimensions it has been shown that weak solutions exist, but not that they are unique.  It has also been shown that if strong solutions exist, they will have to be unique, but it has not been shown that such strong solutions do in fact exist.  The underlying physical issue is whether the velocity field will develop singularities by vortex stretching as it is evolved by the Navier-Stokes equations. 

It  is fortunate that this issue does not arise in numerical simulations because the finiteness of the resolution prevents singularities from developing.  As long as the smallest resolved length scale is smaller by order of magnitudes than the Kolmogorov microscale, the finite resolution approximation of the Navier-Stokes equations models hydrodynamic turbulence quite adequately.  Furthermore, the energy cascade, which is very robust, will not allow any of the Fourier modes to blow out, since all the incoming energy will be transferred to the dissipation range, where it will be disposed of efficiently, given adequate numerical resolution.  Another benefit of the finite resolution model is that the path integrals of the corresponding MSR theory are mathematically rigorous. 

 It should  be noted that the Navier-Stokes equations themselves  are \emph{not} obviously \emph{more} realistic than the finite resolution model because   a ``finite resolution'' \emph{is} imposed on fluid dynamics by Nature  at the point where the existence of discrete molecules is important. Thus, if one introduces the assumption that the finite resolution approximation of the Navier-Stokes equations is a satisfactory physical model all by itself, then one may disregard the mathematical issues associated with the existence, uniqueness, and regularity of the solutions of the Navier-Stokes equations.  This is not an unreasonable assumption \emph{in the inertial range} of three-dimensional turbulence.  We are on less solid ground with respect to the robustness of the cascades of two-dimensional turbulence, but the underlying mathematical issues do not arise in two dimensions. We do not wish to underestimate the importance of the  mathematical issues of existence and uniqueness that remain open for 3D Navier-Stokes; we merely want to highlight the implicit assumption that one makes when one sidesteps these issues, as is done by every  theory published to date.

Another very important issue which is ``hidden under the rug'' is proving the existence of turbulence itself as a consequence of the Navier-Stokes equations.  Unfortunately, the theoretical framework prescribed by the MSR theory cannot account, even in principle, for the existence of the hydrodynamic instability that causes turbulence.  In the MSR framework, it is implicitly assumed that the effect of hydrodynamic instability can be \emph{modeled} by a stochastic forcing term.  The assumption can be justified if one demonstrates that the resulting stochastic behavior of the velocity field in the inertial range is invariant with respect to large-scale perturbations to the statistics of the forcing term.

There is in fact an extension of MSR theory in terms of a supersymmetric path integral that includes two additional fermionic ghost fields \cite{article:Pavlik:1990,article:Gozzi:1988,article:Thacker:1989}.  The surprising result is that correlations involving these additional fields are related to the Lyapunov exponents \cite{article:Reuter:1994} that quantify hydrodynamic instability.  It is therefore possible, in principle, to obtain statistical predictions from this framework with a deterministic forcing as input \cite{article:Thacker:1997}. Whether this is in fact a practical approach remains to be seen.

The assumptions described so far are needed to bring in the machinery of the MSR formalism.  In order to employ the formalism to explain the universality of the direct energy cascade and calculate the intermittency corrections, it is necessary to hypothesize a mathematical description of the energy cascade and use that to narrow down the specific solution which is self-consistent.   Frisch \cite{article:Frisch:1991,book:Frisch:1995} proposed a set of hypothesis consisting of assumptions of statistical symmetry (such as homogeneity, isotropy, self-similarity) and the additional assumption of anomalous dissipation, as an appropriate refinement of Kolmogorov's theory. The nature of the theoretical argument is to show that there is only a unique solution that can be admitted that satisfies the hypothesized statistical symmetries.  A critical review of the assumed statistical symmetries, and local vs. incremental homogeneity in particular,  is part of what concerns us in this paper.

To summarize, we accept the following assumptions on physical or experimental grounds: first, there exists a unique solution to the Navier-Stokes equations that develops hydrodynamic instability for large Reynolds numbers; second, in the limit of fully developed turbulence, incremental homogeneity and incremental isotropy (as defined by Frisch \cite{article:Frisch:1991,book:Frisch:1995}, and see section \ref{sec:kolmdefs}) are reinstated statistically, even if only asymptotically, for the velocity field; third, we accept the hypothesis that there exists an anomalous energy sink at small scales.  These assumptions are a reasonable starting point for analytical theories of turbulence in three dimensions.

\section{Homogeneity and sweeping interactions}

The background on homogeneity is as follows: Taylor, Batchelor, Kraichnan, and others, have been   willing to tolerate the assumption that turbulence is globally homogeneous and isotropic.  However, it was suggested by Kolmogorov himself \cite{article:Kolmogorov:1941} that a far more realistic approach  is to assume local homogeneity and local isotropy.  Both frameworks have been reviewed by Monin and Yaglom \cite{book:Yaglom:1975}.  Kolmogorov also emphasized the importance of studying stationary turbulence, corresponding to the forced-dissipative case, instead of the free decaying case. 

In recent work, Frisch \cite{article:Frisch:1991,book:Frisch:1995} proposed that Kolmogorov's second paper \cite{article:Kolmogorov:1941:1} leads to a reformulation of his theory along three assumptions: first, the assumption of local homogeneity and local isotropy (defined differently than by Kolmogorov, see section \ref{sec:kolmdefs}); second, an assumption of self-similarity; third,  the assumption of an anomalous energy sink.  Using the first and third assumption, according to Frisch, one derives the $4/5$ law from which we obtain $\gz_3=1$.  From the second assumption we have $\gz_n=nh$.  Combined, we obtain the prediction $\gz_n =n/3$.   The assumption of self-similarity, used by Frisch, axiomatically excludes intermittency corrections to the scaling exponents $\gz_n$.    Consequently, the theoretical efforts to calculate the scaling exponents from ``first principles''  essentially aim to weaken this  assumption while tolerating the other two assumptions.  

Some faith in the assumption of an anomalous energy sink, in particular, is based on recent  evidence from numerical simulations \cite{article:Uno:2003} and theoretical evidence from the fusion rules \cite{article:Procaccia:1996:1,article:Procaccia:1996:3}. The assumption of local isotropy can be understood from the principle of linear superposition of the isotropic and anisotropic sectors of the symmetry group $SO (3)$ \cite{article:Procaccia:1999,article:Procaccia:2005}. Finally, the assumption of self-similarity can be understood via $\cZ (h)$ covariance of the statistical theory \cite{article:Procaccia:1998,article:Procaccia:1998:1,article:Procaccia:1998:2}. This leaves then the assumption of local homogeneity.

\subsection{Hierarchical definitions of  homogeneity}

 Let $u_{\ga}(\bfx,t)$ be the Eulerian velocity field, and introduce the Eulerian velocity differences $w_{\ga}$:
\begin{equation}
w_{\ga} (\bfx,\bfxp, t) = u_{\ga}(\bfx,t) - u_{\ga}(\bfxp,t).
\end{equation}
The Eulerian generalized structure function is defined as the ensemble average of the product of such velocity differences
\begin{equation}
F_n^{\ga_1\ga_2\cdots\ga_n}  (\{\bfx, \bfxp\}_n, t) = \avg{\left[ \prod_{k=1}^n w_{\ga_k} (\bfx_k, \bfxp_k,t) \right]},
\end{equation}
where $\{\bfx, \bfxp\}_n$ is shorthand for a list of $n$ position vectors. 

Originally, Frisch \cite{article:Frisch:1991,book:Frisch:1995} wrote his definitions of  local homogeneity, local isotropy, and local stationarity using an ``equivalence in law'' relation.     It should be noted that one should distinguish between \emph{many-time equivalence}, that extends to many-time correlations, and \emph{one-time equivalence} that applies only to one-time correlations.  The clearest way to bring out this distinction is by defining the equivalence relation in terms of  characteristic functionals   defined as
\begin{align}
Z_{\bfw}^{\bfx,\bfxp}[\bfp, t] &= \avg{\exp \left( i \int d\bfx \int d\bfxp \; w_{\ga} (\bfx, \bfxp, t) p_{\ga} (\bfx, \bfxp) )\right) } \\
Z_{\bfw}^{\bfx,\bfxp,t}[\bfp] &= \avg{\exp \left( i \int d\bfx \int d\bfxp \int dt \; w_{\ga} (\bfx, \bfxp, t) p_{\ga} (\bfx, \bfxp,t) )\right) }.
\end{align}
The structure functions can be evaluated from the characteristic functional by variational differentiation and setting $\bfp = \bfzero$.  For example,
\begin{equation}
F_n^{\ga_1\ga_2\cdots\ga_n} (\{\bfx, \bfxp\}_n, t) = \left[ \prod_{k=1}^n \frac{1}{i}\frac{\gd}{\gd p_{\ga_k} (\bfx_k, \bfxp_k) } \right] Z_{\bfw}^{\bfx, \bfxp}[\bfp,t] \Big\vert_{\bfp = 0}.
\end{equation}
The difference between $Z_{\bfw}^{\bfx,\bfxp}[\bfp]$ and $Z_{\bfw}^{\bfx,\bfxp,t}[\bfp]$, is that $Z_{\bfw}^{\bfx,\bfxp}[\bfp]$ contains information only about one-time correlations, whereas $Z_{\bfw}^{\bfx,\bfxp,t}[\bfp]$ contains information about many-time correlations as well.  This is exploited to distinguish between many-time equivalence and one-time equivalence. 

 \begin{definition}
Consider two stochastic fields $v_{\ga} (\bfx, \bfxp, t)$ and $w_{\ga}(\bfx, \bfxp, t)$.  The ``equivalence in law'' relations are defined as
\begin{align}
v_{\ga} (\bfx, \bfxp, t) \seqss w_{\ga} (\bfx, \bfxp, t) &\ifonlyif Z_{\bfv}^{\bfx,\bfxp} [\bfp,t] = Z_{\bfw}^{\bfx,\bfxp} [\bfp,t] \;\forall \bfp \text{ analytic} \\
v_{\ga} (\bfx, \bfxp, t) \seqsst w_{\ga} (\bfx, \bfxp, t) &\ifonlyif Z_{\bfv}^{\bfx,\bfxp,t} [\bfp] = Z_{\bfw}^{\bfx,\bfxp,t} [\bfp] \;\forall \bfp \text{ analytic}.
\end{align}
\end{definition}
Here, $\seqss$ represents one-time equivalence, and $\seqsst$ represents many-time equivalence. Thus, we can  distinguish between one-time global homogeneity $\bfu\in\cH$ and many-time global homogeneity $\bfu\in\cH^{\ast}$:
\begin{align}
\bfu\in\cH &\ifonlyif u_{\ga} (\bfx, t) \seqs  u_{\ga} (\bfx+\bfy, t), \; \forall \bfy\in \bbR^d \\
\bfu\in\cH^{\ast} &\ifonlyif u_{\ga} (\bfx, t) \seqst  u_{\ga} (\bfx+\bfy, t), \; \forall \bfy\in\bbR^d.
\end{align}

A detailed review of previous definitions of \emph{local} homogeneity has been given by Hill \cite{article:Hill:2002:1}.  To discuss   local homogeneity  more carefully, we introduce the following definitions:

\begin{definition}
The velocity field $\bfu$, as a stochastic field, is a member of the homogeneity class $\cH_m (\cA)$ where $\cA \subseteq \bbR^d$ a region in $\bbR^d$, if and only if the ensemble average defined as
\begin{equation}
F_{m,n} \equiv \avg{\left[ \prod_{l=1}^m u_{\ga_l} (\bfx_l,t) \right] \left[ \prod_{k=1}^n w_{\gb_k} (\bfy_k, \bfyp_k,  t) \right]},
\end{equation}
is invariant with respect to a space shift of its arguments $\bfx_l, \bfy_k, \bfyp_k$ for all $n>0$ in the domain $\cA$, i.e.
\begin{equation}
\left( \sum_{l=1}^m \pd_{\ga_l,\bfx_l} + \sum_{k=1}^n (\pd_{\gb_k,\bfy_k} + \pd_{\gb_k,\bfyp_k}) \right) F_{m,n} = 0,\quad \forall \bfx_l, \bfy_k, \bfyp_k \in \cA
\end{equation}
\end{definition}

 \begin{definition}
The velocity field $\bfu$ is a member of the homogeneity class $\cH_m^{\ast}(\cA)$ where $\cA \subseteq \bbR^d$ a region in $\bbR^d$, if and only if the ensemble average defined as
\begin{equation}
F_{m,n}^{\ast} \equiv \avg{\left[ \prod_{l=1}^m u_{\ga} (\bfx_l,t_l) \right] \left[ \prod_{k=1}^n w_{\gb_k} (\bfy_k, \bfyp_k,  t) \right]},
\end{equation}
is invariant with respect to a space shift of its arguments $\bfx_l, \bfy_k, \bfyp_k$ for all $n>0$ in the domain $\cA$, i.e.
\begin{equation}
\left( \sum_{l=1}^m \pd_{\ga_l,\bfx_l} + \sum_{k=1}^n (\pd_{\gb_k,\bfy_k} + \pd_{\gb_k,\bfyp_k}) \right) F_{m,n}^{\ast} = 0,\quad \forall \bfx_l, \bfy_k, \bfyp_k \in \cA
\end{equation}
\end{definition}
We also write $\cH_m \equiv \cH_m (\bbR^d)$ and  $\cH_m^{\ast} \equiv \cH_m^{\ast}  (\bbR^d)$. The distinction between $\cH_m(\cA)$ and $\cH_m^{\ast}(\cA)$ is that the former requires translational invariance on the one-time correlation tensor $F_{m,n}$, whereas the latter requires translational invariance on the many-time correlation tensor $F_{m,n}^{\ast}$, both over the domain $\cA$. 

We also define the following transfinite homogeneity classes:
\begin{equation}
\cH_{\gw}(\cA) = \bigcap_{k \in \bbN} \cH_k (\cA)
\quad\text{ and }\quad
\cH_{\gw}^{\ast}(\cA) = \bigcap_{k \in \bbN} \cH_k^{\ast}(\cA).
\end{equation}
In these homogeneity classes  the ensemble average of any product of velocities multiplied with any product of velocity differences will be invariant under spatial shifting.  Note that even this homogeneity class is weaker than \emph{global homogeneity}. We will also distinguish between \emph{one-time global homogeneity} $\bfu\in\cH$ and \emph{many-time global homogeneity} $\bfu\in\cH^{\ast}$, which are defined as
 \begin{align}
 \bfu\in\cH &\ifonlyif u_{\ga} (\bfx, t) \seqs  u_{\ga} (\bfx+\bfy, t), \; \forall \bfy\in \bbR^d \\
 \bfu\in\cH^{\ast} &\ifonlyif u_{\ga} (\bfx, t) \seqst  u_{\ga} (\bfx+\bfy, t), \; \forall \bfy\in\bbR^d.
 \end{align}

\begin{remark}
An immediate consequence of these definitions is that the homogeneity classes are hierarchically ordered, according to the following relations
\begin{align}
\cH &\subseteq \cH_{\gw} (\cA)\subseteq \cH_k(\cA), \; \forall k \in \bbN, \\
\cH^{\ast}&\subseteq \cH_{\gw}^{\ast} (\cA)\subseteq \cH_k^{\ast}(\cA), \; \forall k \in \bbN, \\
\cH_a (\cA)&\subseteq \cH_b(\cA) \land \cH_a^{\ast} (\cA)\subseteq \cH_b^{\ast} (\cA),\;\forall a, b \in \bbN : a > b, \\
\cH_a^{\ast} (\cA)& \subseteq \cH_a (\cA), \;\forall a \in \bbN.
\end{align}
\end{remark}

\subsection{Remarks on Kolmogorov's and Frisch's definition of local homogeneity}
\label{sec:kolmdefs}

The term ``local homogeneity'' is usually identified with the definition that was given by Kolmogorov \cite{article:Kolmogorov:1941}. However, in his reformulation of the Kolmogorov 1941 theory, Frisch \cite{article:Frisch:1991,book:Frisch:1995} identified local homogeneity, local isotropy, and local stationarity with incremental homogeneity, incremental isotropy, and incremental stationarity. The definitions that he gave read:
\begin{align*}
&\text{Locally stationary: } & w_{\ga} (\bfx, \bfxp, t) &\seqss w_{\ga} (\bfx, \bfxp, t+\gD t) \; ,\forall \gD t \in \bbR.\\
&\text{Locally homogeneous: }& w_{\ga} (\bfx, \bfxp, t) &\seqss w_{\ga} (\bfx + \bfy, \bfxp + \bfy, t) \; ,\forall \bfy \in \bbR^d.\\
&\text{Locally isotropic: }& w_{\ga} (\bfx, \bfxp, t) &\seqss w_{\ga} (\bfx_0 + A (\bfx-\bfx_0), \bfx_0 + A (\bfxp-\bfx_0), t) \; ,\forall A \in SO (d).\\
\end{align*}
 Using our notation, the condition of incremental homogeneity can be written as $\bfu \in \cH_0(\cA)$. It should be stressed that Frisch postulated that these symmetries  are valid \emph{asymptotically} for space shifts and time shifts up to a relevant order of magnitude and proposed them  as reasonable hypotheses to be used as the basis for a modern reformulation of Kolmogorov's 1941 theory \cite{article:Frisch:1991,book:Frisch:1995}.

 To motivate his hypotheses, Frisch argues that homogeneity, isotropy, and time invariance are satisfied by the Navier-Stokes equations and they are violated only by the boundary conditions or any other relevant means of generating turbulence.  However, he suggests that for high Reynolds numbers, when the turbulent motion is governed by a strange attractor, the symmetries of the governing equation are restored asymptotically for small scales.  Velocity differences are used to localize the symmetry to small scales.

The paradox inherent in this argument is that we cannot write governing equations for the velocity differences, exclusively in terms of velocity differences.  A nonlinear term involving the velocity field, representing the sweeping interactions, is inevitable.  As we shall argue below, the stronger homogeneity assumption $\bfu \in \cH_1 (\cA)$ is required to drop this term. Furthermore, we will argue that $\bfu \in \cH_1 (\cA)$ is also required to derive the $4/5$-law, which is the first step in Frisch's argument. Similar concerns were raised recently by Frisch \cite{article:Frisch:2005} who questioned the self-consistency of local homogeneity, both in the sense of incremental homogeneity and  in the sense of Kolmogorov.

As for Kolmogorov, in his first paper \cite{article:Kolmogorov:1941}, he defined local homogeneity in a very interesting way. Instead of using the Eulerian velocity differences $w_{\ga} (\bfx, \bfxp, t)$, he used the following quantity:
\begin{align}
\bfY (\bfx_0, t_0 | \bfx, t) &= \bfx - \bfx_0 - (t-t_0) \bfu (\bfx_0, t_0) \\
\bfw_{Kol}  (\bfx_0, t_0 | \bfx, t) &= \bfu (\bfY  (\bfx_0, t_0 | \bfx, t), t) - \bfu (\bfx_0, t_0).
\end{align}
Here, $\bfY$ represents the approximate displacement of a fluid particle that is being used as a frame of reference. Because of its dependence on the velocity field, it is itself a stochastic variable. Kolmogorov employed the probability density function of $\bfw_{Kol}$ in his definitions.  Furthermore, he included the requirement of local stationarity in his definition of local homogeneity. As will become apparent in  section 4, Kolmogorov's representation of velocity differences is in fact a precursor of the quasi-Lagrangian representation, and we shall call it the \emph{Kolmogorov quasi-Lagrangian representation}.  Although Kolmogorov does not discuss explicitly the problem of sweeping interactions, it is interesting that he foresaw to this extent the need for an non-Eulerian representation of the velocity field.

Another curious feature of the Kolmogorov definition is that it appears to use a conditional ensemble average conditioned on the statement $\bfu (\bfx_0, t_0) = \bfv$ instead of the usual unconditional ensemble average,  and includes independence with respect to $\bfv$ as part of his definition of local homogeneity. Equivalently,  one may use a conditional average conditioned on the location of the fluid particle $\bfY (\bfx_0, t_0 | \bfx, t)= \bfy$ and assume independence with respect to $\bfy$. The equivalence depends on using Kolmogorov's  quasi-Lagrangian transformation, and  is not applicable if one replaces it with the Belinicher-L'vov   quasi-Lagrangian transformation.  

For the case $t=t_0$ it is easy to see that the Kolmogorov definition is as strong as $\bfu\in\cH_0$. However, more generally, the velocity differences used by Kolmogorov are evaluated at two different times  $t$ and $t_0$. For this reason, I find it very unlikely that Kolmogorov's definition can be shown to be as strong as $\bfu\in\cH_1$. On the other hand, the use of the conditional ensemble average and the assumption that that average is independent of $\bfv$ probably strengthens the definition in unforeseen ways and may have some interesting consequences.  In section 4.4 we show that a modified version of the Kolmogorov definition of local homogeneity is  equivalent or stronger than $\cH_1$, thus providing the assumptions needed to prove the $4/5$-law and eliminate the sweeping interactions.

A detailed discussion and criticism of Kolmogorov's definition of local homogeneity is  given by Frisch \cite{article:Frisch:2005}. In the same paper, contrary to his prior work \cite{article:Frisch:1991,book:Frisch:1995}, Frisch distinguishes the term local homogeneity from incremental homogeneity, and assigns Kolmogorov's definition as the definition of local homogeneity.  We have seen that the definition of local homogeneity by Kolmogorov includes an assumption of incremental stationarity and also an assumption of a type of random Galilean invariance (i.e. independence with respect to $\bfv$).  Incremental stationarity can be true even when incremental homogeneity is not true. Furthermore, as I shall argue in section 5 of this paper, we should intend to derive random Galilean  invariance from the theory instead of assuming it. Consequently, the original definitions \cite{article:Frisch:1991,book:Frisch:1995} of Frisch have the practical advantage of conveniently separating these assumptions from each other, and the conceptual advantage of not assuming too much.

\subsection{Balance equations and sweeping}
\label{sec:balance}

The clearest way to analyze the effect of the sweeping interactions on the theory of hydrodynamic turbulence is by employing the balance equations of the Eulerian generalized structure functions.  These balance equations were introduced by L'vov and Procaccia \cite{article:Procaccia:1996:3} in a landmark paper, and they are derived as follows.

The Navier-Stokes equations, where the pressure term has been eliminated, read
\begin{equation}
\pderiv{u_{\ga}}{t} + \cP_{\ga\gb} \partial_{\gc} (u_{\gb}u_{\gc}) = \nu \del^2 u_{\ga} + \cP_{\ga\gb} f_{\gb},
\end{equation}
where $\cP_{\ga\gb}$ is the projection operator defined as
\begin{equation}
\cP_{\ga\gb} = \gd_{\ga\gb} - \partial_{\ga}\partial_{\gb} \del^{-2},
\end{equation}
and $\partial_{\ga}$ represents spatial differentiation with respect to $x_{\ga}$.  Repeated indices imply summation of components.  The balance equations are obtained by differentiating the definition of $F_n$ with respect to time $t$ and substituting the Navier-Stokes equations.  This leads to exact  equations of the form
\begin{equation}
\pderiv{F_n}{t} + D_n = \nu J_n + Q_n,
\label{eq:balance}
\end{equation}
where $D_n$ represents the contributions from the nonlinear term, $J_n$ the contributions of the dissipation term, and $Q_n$ the contribution from the forcing term. To write the terms concisely, we use the following abbreviations to represent aggregates of arguments:
\begin{equation}
\begin{split}
\bfX &= (\bfx , \bfxp) \\ 
\{\bfX \}_n &= \{\bfX_{1} , \bfX_{2} , \ldots , \bfX_{n}\}\\ 
\{\bfX_\ga\}_n^k &= \{\bfX_{1} , \ldots , \bfX_{k-1} , \bfX_{k+1} , \ldots , \bfX_{n}\}.
\end{split}
\end{equation} 
The terms themselves read as follows. The forcing contribution is given by
\begin{equation}
Q_n^{\ga_1\ga_2\cdots\ga_n} (\{\bfX\}_n, t) = \sum_{k=1}^n \avg{\left[ \prod_{l=1,l\neq k}^n w_{\ga_l}(\bfx_l, \bfxp_l,t) \right]\cP_{\ga_k\gb}(f_{\gb}(\bfx_k,t) - f_{\gb}(\bfxp_k,t))}.
\end{equation}
The dissipation term is given by
\begin{equation}
J_n^{\ga_1\ga_2\cdots\ga_n} (\{\bfX\}_n, t) = \cD_n F_n^{\ga_1\ga_2\cdots\ga_n} (\{\bfX\}_n, t)= \sum_{k=1}^n  (\nabla^{2}_{\bfx_k} + \nabla^{2}_{\bfxp_k}) F_n^{\ga_1\ga_2\cdots\ga_n}(\{\bfX\}_n, t),
\end{equation}
where $\nabla^{2}_{\bfx_k}$ differentiates with respect to $\bfx_k$, and similarly  $\nabla^{2}_{\bfxp_k}$ differentiates with respect to $\bfxp_k$.  

The remarkable result, shown in \cite{article:Procaccia:1996:3}, is that the term $D_n$ that represents the contribution of the nonlinear term can be rewritten as 
$D_n = \cO_n F_{n+1} + I_n$
where $\cO_n$ is a linear integrodifferential operator with general form
\begin{align}
\cO_n F_{n+1} &= \sum_{k=1}^n \cO_{nk} F_{n+1}\\
\cO_{nk} F_{n+1} &= \int \cO (\bfX_k ,\bfY_1 ,\bfY_2) \;F_{n+1} (\{\bfX\}^k_n, ,\bfY_1 ,\bfY_2,t)\; d\bfY_1 d\bfY_2,
\end{align}
and $I_n$ is given by 
\begin{equation}
I_n^{\ga_1\ga_2\cdots\ga_n} (\{\bfX\}_n, t) = \sum_{k=1}^n (\partial_{\gb, \bfx_{k}} +\partial_{\gb, \bfxp_{k}} )\left\langle \cU_\gb (\{\bfX\}_n, t) \left[  \prod_{l=1}^n w_{\ga_l} (\bfX_{l} , t) \right] \right\rangle, 
\end{equation}
where $\cU_\gb (\{\bfX\}_n, t)$ is defined as
\begin{equation}
\cU_\ga (\{\bfX\}_n, t) = \frac{1}{2n}\sum_{k=1}^n \left (u_{\alpha}(\bfx_{k}, t)+  u_{\alpha}(\bfxp_{k} ,t) \right ).
\end{equation}
The first term,  $\cO_n F_{n+1}$, includes the effect of pressure and part of the advection term, and the detailed form of the operator $\cO_n$  has been given in Ref  \cite{lfthesis:2006}. The second term, $I_n$, represents exclusively the effect of the sweeping interactions. 

This decomposition makes rigorous the notion that the nonlinear interactions in the Navier-Stokes equations consist of local interactions that are responsible for the energy cascade and sweeping interactions which would disrupt the energy cascade if they contaminated the inertial range.  It also exposes the conditions under which the sweeping interactions can be neglected.  We learn that if the ensemble average of the velocity product that appears in the definition of $I_n$ is invariant under a spatial shift, then the derivatives of that ensemble average will add up to zero.  And here lies the heart of the problem. The assumption $\bfu\in\cH_0$ by itself is not sufficient to set $I_n =0$.  Global homogeneity $\bfu\in\cH$ is sufficient, but it is a stronger assumption than what is required. 

\begin{remark}
The condition of incremental homogeneity, is written as $\bfu \in \cH_0(\cA)$.  The homogeneity condition needed to eliminate the sweeping interactions over the domain $\cA$  is  $\bfu \in \cH_1(\cA)$.
\end{remark}

It should be noted that the local term  $\cO_n F_{n+1}$ and the dissipation term  $J_n$ preserve the incremental homogeneity condition $\bfu \in \cH_0(\cA)$. The two terms in the balance equations that can potentially violate incremental homogeneity,  are the sweeping term  $I_n$ and the forcing term  $Q_n$. Asymptotic incremental homogeneity cannot be disrupted in the inertial range by the forcing term \emph{if} the forcing spectrum is confined to large scales. The uncontrolled quantity is the sweeping term  $I_n$. Recently, Frisch \cite{article:Frisch:2005} questioned the consistency of local homogeneity, in the sense of Kolmogorov, and incremental homogeneity as a framework for studying hydrodynamic turbulence. We see that incremental homogeneity can be a consistent framework on the condition that the sweeping term  $I_n$ is dominant only at large scales with its influence forgotten as the energy cascades to smaller scales. If that is the case, then none of the other terms in the balance equations violate incremental homogeneity. This is discussed in further detail in section 5.

\subsection{Remarks on the $4/5$-law proof}

In his second paper, Kolmogorov \cite{article:Kolmogorov:1941:1} employed an argument that is distinct from dimensional analysis to explain the claim that $\gz_n = n/3$.  He derived the $4/5$-law from which he obtained $\gz_3=1$, and used a scaling assumption to obtain $\gz_2=2/3$.  Frisch's \cite{article:Frisch:1991,book:Frisch:1995} contribution was his observation that the scaling argument can be extended to account for all the scaling exponents $\gz_n$.  With this extension, Kolmogorov's second paper \cite{article:Kolmogorov:1941:1} is then an equivalent reformulation of the dimensional analysis argument of his first paper \cite{article:Kolmogorov:1941}.  The superiority of the extended argument is that at least one of the scaling exponents is established rigorously. One also bypasses the universality criticism, appearently attributed to Landau, of the original similarity hypothesis of Kolmogorov. In his book, Frisch \cite{book:Frisch:1995} gave a more detailed account of his argument, but he didn't derive the $4/5$-law on the basis of incremental homogeneity and incremental isotropy as prescribed by his framework; he used instead global homogeneity and global isotropy. The same holds for the alternative proof by Rasmussen \cite{article:Rasmussen:1999}. An old proof by Monin \cite{article:Monin:1959}  and Monin and Yaglom \cite{book:Yaglom:1975} claimed to prove the $4/5$-law  on the basis of local homogeneity and local isotropy, but it was criticized by Lindborg \cite{article:Lindborg:1996}.  The criticism was addressed by Hill \cite{article:Hill:1997} who gave a corrected proof. 

In particular, the criticism of Lindborg \cite{article:Lindborg:1996} was that it was not proved that the correlations involving the pressure field gradient and the velocity field can be eliminated on the basis of local isotropy from the equation that governs the time derivative of the second order structure function tensor. Hill \cite{article:Hill:1997} resolved this objection by supplying the needed proof. The principle behind the proof is reflected, in a wider sense, by the mathematical form of the general sweeping term $I_n$ where there is only a local differential operator. The elimination of the nonlocal integral operator from $I_n$ represents the elimination of any contributions by the pressure gradient term to $I_n$ that would break incremental homogeneity. The pressure gradient \emph{does} contribute to the term $\cO_n F_{n+1}$  a non-local integrodifferential operator.  However, because $\cO_n F_{n+1}$ can be expressed exclusively in terms of the velocity differences, it preserves incremental homogeneity.  

Nevertheless, the proof by Monin and Yaglom \cite{book:Yaglom:1975}, as far as our intentions are concerned, has an additional shortcoming, which has also been noticed independently by Frisch \cite{article:Frisch:2005}: it concerns the elimination of the terms associated with the sweeping interactions.   If we refer to the part of the discussion leading to  equation (22.14) of  Monin and Yaglom \cite{book:Yaglom:1975}, we learn that \emph{they are  using the Belinicher-L'vov quasi-Lagrangian transformation to eliminate the sweeping interaction term}! This can be made more clear if the reader compares the argument involving the two unnumbered equations that precede equation (22.14) of  Monin and Yaglom \cite{book:Yaglom:1975} with section 4 and appendix \ref{sec:qlgoveqs}. The intention of this argument, according to Monin and Yaglom \cite{book:Yaglom:1975}, is to ``... transform the Navier-Stokes equations so that they contain only the velocity differences and their derivatives''. This is precisely what the quasi-Lagrangian formulation does. As we shall argue in the next section of this paper, applying the inverse transformation back to the Eulerian representation uses an assumption of homogeneity stronger than incremental homogeneity, but this time $\bfu \in \cH_{\gw}^{\ast}$.

As far as the theory of the scaling exponents is concerned, it is only necessary to know $\gz_3$.  An elegant way to calculate $\gz_3$ is from the solvability condition of the homogeneous equation $\cO_2 F_3 =0$ \cite{article:Procaccia:1996:1,article:Procaccia:1999}. The idea here is to use the conservation of energy to show that
\begin{equation}
\begin{split}
\cO_2 F_3 (\bfx_1, \bfxp_1, \bfx_2, \bfxp_2 ) &= \frac{1}{2}\frac{d[S_3 (r_{12}) - S_3 (r_{12'})]}{dr_1} +  \frac{1}{2}\frac{d[S_3 (r_{1'2'}) - S_3 (r_{1'2})]}{dr_{1'}} \\
&= A [r_{12}^{\gz_3-1} - r_{12'}^{\gz_3-1} + r_{1'2'}^{\gz_3-1} - r_{1'2}^{\gz_3-1}].
\label{eq:o2f3}
\end{split}
\end{equation}
where $r_{12} = \nrm{\bfx_1 - \bfx_2}$, etc. It follows that the equation $\cO_2 F_3 =0$ will be satisfied for any configuration of velocity differences if and only if $\gz_3=1$.  The homogeneous equation can be obtained from the balance equations in the limit of infinite Reynolds number.  For the case of finite Reynolds number, there is a homogeneous and particular solution to the generalized structure functions that are linearly superimposed \cite{submitted:Tung:2003,submitted:Tung:2003:1}.   Then the calculation of $\gz_3$ is relevant only for the homogeneous solution. Aside from this issue, this argument too requires that we  set $I_2=0$. Dropping $I_2$ cannot be justified under incremental homogeneity, in the sense of $\bfu\in\cH_0$, and it requires the condition $\bfu\in\cH_1(\cA)$.  We arrive then to the following conclusion.

\begin{remark}
 The homogeneity condition needed to  establish $\gz_3=1$ over the domain $\cA$  is  $\bfu \in \cH_1(\cA)$.
\end{remark}

It should be noted that even though Hill \cite{article:Hill:1997} has claimed to show the $4/5$-law on the basis of local homogeneity and local isotropy, his definition of local homogeneity is mathematically stronger than the definition  $\bfu\in\cH_0(\cA)$ used in the Frisch framework, and it is in fact very similar to $\bfu\in\cH_1(\cA)$ (also see section 4.1 of \cite{article:Hill:2002}). Consequently, while his proof correctly follows from his stated assumptions, it cannot be used from within the Frisch framework of hypotheses to prove the $4/5$-law without invoking additional assumptions.

It is possible to derive a rigorous version of the $4/5$-law that does not require assumptions of homogeneity, isotropy, stationarity, and not even an ensemble average \cite{article:Robert:2000,article:Eyink:2003:1,article:Eyink:2003:2}.  This is done by rephrasing the statement to be proven.  Specifically, it has been shown that
\begin{equation}
\lim_{\gD t\goto 0}\lim_{r \goto 0}
\lim_{\nu\goto 0} \int_t^{t+\gD t} d\tau \int_{SO(3)}\frac{d\gW (A)}{4\pi}\int_{\cB}\frac{d\bfx }{V (\cB)} \frac{S_3 (\bfx, rA\bfe)}{r} = -\frac{4}{5}\gee_{\cB},
\end{equation}
for almost every (Lebesgue) point $t$ in time, where $\bfe$ is a unit vector, $\cB \subseteq \bbT^3$ is  a local region in a periodic boundary domain $\bbT^3$ (topologically equivalent to a torus) with volume $V(\cB)$, and $\gee_{\cB}$ is the local dissipation rate over the region $\cB$ given by
\begin{equation}
\gee_{\cB} \equiv \lim_{\nu\goto 0}\frac{1}{V(\cB)}\int d\bfx\, \gee (\bfx, t),
\end{equation}
where $\gee (\bfx,t) = (1/2) \nu \avg{s_{\ga\gb}(\bfx,t) s_{\ga\gb}(\bfx,t)}$ is the dissipation rate density at $(\bfx,t )$ and $s_{\ga\gb}\equiv \pda u_{\gb} + \pdb u_{\ga}$ is the local strain tensor. A similar result was obtained earlier  by Nie and Tanveer \cite{article:Tanveer:1999}.

It should be noted that this result does not contradict our previous remark.  Although the need to make assumptions appears to have been eliminated, this is done so at the price of proving a statement that is mathematically weaker. In  the original formulation of the $4/5$-law, aside from an ensemble average, all the integrals are absent. These integrals represent an interesting way of obviating the symmetry assumptions needed to prove the $4/5$-law in its original formulation.

Recently, there has been considerable interest in extending the $4/5$-law to account for deviations from the theoretical prediction caused by the violation of incremental isotropy \cite{article:Lindborg:1999:1,article:Antonia:2001,article:Antonia:2002,article:Zhou:2004,article:Burattini:2004}. From the viewpoint of the experimentalist these extensions make it possible to confirm the validity of the $4/5$-law against experimental data. From the viewpoint of the theorist, deviations from incremental isotropy can be accounted for with the $SO(3)$ group decomposition method  \cite{article:Procaccia:1999,article:Procaccia:2005}.

\section{The quasi-Lagrangian formulation}

The essence of the  quasi-Lagrangian formulation (also called the \emph{Belinicher-L'vov transformation})  is to look at turbulence using a fluid particle as a non-inertial frame of reference.  The representation is Lagrangian because we involve fluid particles, but it is not completely Lagrangian because the fluid particle trajectory is only used to define a new frame of reference, and we continue to look at the velocity field in an  Eulerian manner.  It is understood, of course, that the only interesting statistics are those involving points within a sphere centered on the moving fluid particle with radius on the order of the integral length scale $\ell_0$.

Let $u_{\ga}(\bfx,t)$ be the Eulerian velocity field, and let  $\rho_{\alpha}(\bfx_0 ,t_0 |t)$ be the position of the unique fluid particle initiated at $(\bfx_0 ,t_0)$ at time $t$ relative to its initial position at time $t_0$. The transformation is done in two steps. First, we introduce $v_{\alpha}(\bfx_0 ,t_0 | \bfx,t)$ as the Eulerian velocity with respect to the original inertial frame of reference with a space shift that follows the fluid particle:
\begin{equation}
\begin{split}
\rho_{\alpha}(\bfx_0 ,t_0 |t) &= \int_{t_0}^{t} d\tau \; u_{\alpha} (\bfx_0 +
\rho (\bfx_0 ,t_0 |\tau) , \tau) \\ v_{\alpha}(\bfx_0 ,t_0 | \bfx,t) &=
u_{\alpha} (\bfx + \rho (\bfx_0 ,t_0 |t) , t).\\
\end{split}
\end{equation}
Then, to complete the transformation we must subtract the velocity of
the fluid particle uniformly, so that the particle itself will appear to be motionless:
\begin{equation}
\begin{split}
w_{\alpha}(\bfx_0 ,t_0 | \bfx,t) &= v_{\alpha}(\bfx_0 ,t_0 | \bfx,t) - \pD{t}
\rho_{\alpha}(\bfx_0 ,t_0 |t) = v_{\alpha}(\bfx_0 ,t_0 | \bfx,t) -
v_{\alpha}(\bfx_0 ,t_0 |\bfx_0 ,t) \\ &= u_{\alpha} (\bfx + \rho (\bfx_0 ,t_0 |t)
, t) - u_{\alpha} (\bfx_0 + \rho (\bfx_0 ,t_0 |t) , t).
\end{split}
\end{equation}
We define $w_{\alpha}(\bfx_0 ,t_0 | \bfx, t)$ as the quasi-Lagrangian velocity field, and introduce the quasi-Lagrangian velocity difference $W_\ga (\bfx_0, t_0 | \bfx, \bfxp, t)$  given by
\begin{align}
W_\ga (\bfx_0, t_0 | \bfx, \bfxp, t) &\equiv w_\ga (\bfx_0, t_0 | \bfx, t) - w_\ga (\bfx_0, t_0 | \bfxp, t) = v_\ga (\bfx_0, t_0 | \bfx, t) - v_\ga (\bfx_0, t_0 | \bfxp, t).
\end{align}
Differentiating with respect to time, and substituting the Navier-Stokes equations, gives an equation of the form
\begin{equation}
\pderiv{W_\ga}{t} + \cV_{\ga\gb\gc} W_{\gb} W_{\gc} = \nu (\lapl_\bfx + \lapl_\bfxp) W_\ga + F_\ga,
\label{eq:qlns}
\end{equation}
where $F_\ga (\bfx_0, t_0 | \bfx, \bfxp, t)$ is the quasi-lagrangian forcing, and $\cV_{\ga\gb\gc}$ is a bilinear integrodifferential operator of the form
\begin{equation}
\cV_{\ga\gb\gc} W_{\gb} W_{\gc} \equiv \iint d\bfX_\gb d\bfX_\gc \, V_{\ga\gb\gc} (\bfx_0 | \bfX_\ga,\bfX_\gb,\bfX_\gc) W_\gb (\bfX_\gb) W_\gc (\bfX_\gc),
\end{equation}
with $ V (\bfx_0 | \bfX_\ga,\bfX_\gb,\bfX_\gc) $ the corresponding kernel (see appendix \ref{sec:qlgoveqs} for more details). The remarkable feature of this equation is that all the terms, and most especially the nonlinear term, are written in terms of velocity differences.  Fundamentally, this is the reason why the quasi-Lagrangian transformation eliminates the sweeping interactions and renormalizes the MSR diagrammatic theory.

The key issue is whether it is possible to switch back to the Eulerian representation without reintroducing the sweeping interactions.  In a short appendix to their paper, L'vov and Procaccia \cite{article:Procaccia:1995:1} showed that \emph{in stationary turbulence the ensemble average of the same time   quasi-Lagrangian velocity differences is equal to the ensemble   average of the corresponding Eulerian velocity differences.} The same appendix is also found in a previous  unpublished paper \cite{article:Lebedev:1994:2}. The proof requires stationarity of the Eulerian velocity field, and incompressibility.  A homogeneity condition is also used, which is described as ``translational invariance''.

In this section, we would like to carefully re-examine this proof, the assumptions needed to make it work, and the relationship between this result and other claims that one might reasonably make about the quasi-Lagrangian velocity differences.  Part of our motivation is the crucial importance of this result; the entire L'vov-Procaccia theory \cite{article:Procaccia:1995:1,article:Procaccia:1995:2,article:Procaccia:1996,article:Procaccia:1997,article:Procaccia:1998,article:Procaccia:1998:1,article:Procaccia:1998:2,article:Procaccia:2000} stands or falls on the validity of this argument. As we have discussed previously in section 2, the L'vov-Procaccia theory and the Giles theory \cite{article:Giles:2001} are the only two  theories that can explain mathematically the reason why the inertial range of three-dimensional turbulence has intermittency corrections.  Our main interest is to show that the proof requires that we assume $\bfu \in\cH_{\gw}^{\ast}$, which is a stronger condition than what is actually needed to eliminate the sweeping interactions or to prove the $4/5$-law  ($\bfu \in \cH_1 (\cA)$). Preliminaries are given in section 4.1 and section 4.2, and the proof itself is discussed in section 4.3.

\subsection{Characterizations of the claim}

Let $\cF_n (\bfx_0, t_0 | \{ \bfx,\bfxp\}_n, t)$ be the generalized structure function in the quasi-Lagrangian representation, defined as
\begin{equation}
\cF_n (\bfx_0, t_0 | \{ \bfx,\bfxp\}_n, t) = \avg{\left[ \prod_{k=1}^n W_{\ga_k} (\bfx_0, t_0 | \bfx_k,\bfxp_k, t) \right]}.
\end{equation}
The claim of L'vov and Procaccia \cite{article:Procaccia:1995:1} was that  it can be shown that
\begin{equation}
\cF_n (\bfx_0, t_0 | \{ \bfx,\bfxp\}_n, t) = F_n (\{ \bfx,\bfxp\}_n, t), \forall n\in\bbNs
\end{equation}
which can be rewritten equivalently as
\begin{equation}
W_\ga (\bfx_0, t_0 | \bfx, \bfxp, t) \seqss w_\ga (\bfx,\bfxp,t).
\label{eq:theclaim}
\end{equation}
As a first step, consider the following easy-to-prove propositions which give equivalent characterizations of the claim \eqref{eq:theclaim}:

\begin{proposition}
\label{prop:stat}
The claim \eqref{eq:theclaim} holds if and only if the quasi-Langrangian velocity is incrementally stationary with respect to $t_0$:
\begin{equation}
W_\ga (\bfx_0, t_0+ \gD t | \bfx, \bfxp, t) \seqss W_\ga (\bfx_0, t_0 | \bfx, \bfxp, t ), \; \forall \gD t \in \bbR - \{0\} 
\label{eq:qlstat}
\end{equation} 
\end{proposition}

 \begin{proof}
 ($\Rightarrow$): Assume that the claim  \eqref{eq:theclaim} holds. Then, it follows that
\begin{equation}
 W_\ga (\bfx_0, t_0 + \gD t| \bfx, \bfxp, t) \seqss w_\ga (\bfx, \bfxp, t) \seqss W_\ga (\bfx_0, t_0 | \bfx, \bfxp, t).
\end{equation}
 ($\Leftarrow$): Now assume that the quasi-Lagrangian velocity field is incrementally stationary with respect to $t_0$. Using the evaluation $ \lim_{t_0\goto t} \gr_\ga (\bfx_0, t_0 | t) = 0$, it follows that
\begin{align}
 W_\ga (\bfx_0, t_0 | \bfx, \bfxp, t) &\seqss \lim_{t_0\goto t} W_\ga (\bfx_0, t_0 | \bfx, \bfxp, t) \seqss \lim_{t_0\goto t} w_\ga ( \{\bfx, \bfxp \} + \bfgr (\bfx_0, t_0 | t), t) \\
 &\seqss w_\ga ( \bfx  , \bfxp , t),
\end{align}
 \end{proof}

 \begin{proposition}
 \label{prop:rstat}
 Assume incremental stationarity on the Eulerian velocity field: 
 \begin{equation}
 w_\ga (\bfx, \bfxp, t) \seqss w_\ga (\bfx, \bfxp, t + \gD t), \; \forall \gD t \in \bbR. 
 \label{eq:eulstat}
 \end{equation}
 Then, the claim \eqref{eq:theclaim} holds if and only if the quasi-Langrangian velocity is incrementally stationary with respect to $t$:
 \begin{equation}
 W_\ga (\bfx_0, t_0 | \bfx, \bfxp, t) \seqs W_\ga (\bfx_0, t_0 | \bfx, \bfxp, t + \gD t), \; \forall \gD t \in \bbR - \{0\} 
 \end{equation} 
 \end{proposition}

  \begin{proof}
  ($\Rightarrow$): Assume that the claim  \eqref{eq:theclaim} holds. Then,
  \begin{align*}
  W_\ga (\bfx_0, t_0 | \bfx, \bfxp, t + \gD t), &\seqss w_\ga (\bfx, \bfxp, t + \gD t) \seqss w_\ga (\bfx, \bfxp, t) \seqss W_\ga (\bfx_0, t_0 | \bfx, \bfxp, t).
  \end{align*}
  ($\Leftarrow$): Now assume that the quasi-Lagrangian velocity field is incrementally stationary  with respect to $t_0$. Using the evaluation $ \lim_{t_0\goto t} \gr_\ga (\bfx_0, t_0 | t) = 0$, it follows that,
  \begin{align*}
  W_\ga (\bfx_0, t_0 | \bfx, \bfxp, t) &\seqss \lim_{t\goto t_0} W_\ga (\bfx_0, t_0 | \bfx, \bfxp, t)\seqss \lim_{t\goto t_0} w_\ga ( \{\bfx, \bfxp \} + \bfgr (\bfx_0, t_0 | t), t)  \\
  &= w_\ga ( \bfx  , \bfxp , t_0)\seqss w_\ga ( \bfx  , \bfxp , t).
  \end{align*}
  \end{proof}

The implication of these propositions is that the relationship between the Eulerian and the quasi-Lagrangian formulations can be established as an immediate consequence of stationarity of the quasi-Lagrangian velocity field with respect to $t_0$.  As Lebedev and L'vov  \cite{article:Lebedev:1994:2} noted, the variable $t_0$ does not appear anywhere in the quasi-Lagrangian Navier-Stokes equations \eqref{eq:qlns}, consequently \emph{the form} of the governing equations allows stationary solutions with respect to $t_0$.  However, to assert that the quasi-Lagrangian velocity field \emph{is} stationary, it is necessary to assume that the quasi-Lagrangian forcing field is also stationary.  Since the definition of the quasi-Lagrangian forcing field entangles the Eulerian forcing field $f_\ga$ with the trajectory field $\gr_\ga$, and since the trajectory field itself is not time invariant (due to the initial condition $\rho_{\alpha}(\bfx_0 ,t_0 |t_0 ) =0$), we cannot make this assumption without justification.  This was the reason, cited by Lebedev and L'vov  \cite{article:Lebedev:1994:2}, for the rigorous proof which is the topic of this section. Proposition \ref{prop:rstat} shows that assuming stationarity in the quasi-Lagrangian representation is sufficient to prove the claim \eqref{eq:theclaim}, and thus this assumption implicitly introduces in the Eulerian frame the conditions needed to prove the claim.

\subsection{MSR theory for Lagrangian trajectories}

  The governing equation for $\gr_{\ga}$ is 
\begin{equation}
\pderiv{\rho_{\ga}(\bfx_0, t_0 | t)}{t} = u_{\ga}(\bfx_0 + \rho (\bfx_0, t_0 |
t), t),
\end{equation}
 with initial condition $\rho_{\alpha}(\bfx_0 ,t_0 |t_0 ) =0$.  Deriving the stationarity condition \eqref{eq:qlstat}  requires an MSR theory where the velocity field $u_{\ga}$ can be thought of as the forcing field with known statistical properties, and the Lagrangian trajectories field $\gr_{\ga}$ as the governed field whose properties we wish to deduce. Unfortunately we may not apply the standard MSR theory   because the equation itself does not assume the standard form $N_{\ga}[\gr]=u_{\ga}$ with $u_\ga$ independent of $\gr$ and furthermore the initial condition is set at a finite time $t_0$ and not at $t_0\rightarrow -\infty $. We need to develop the statistical theory from scratch, and for that purpose the path integral formulation is most expedient.

Note that every value of $t_0$ corresponds to a distinct initial value problem.  We may therefore treat, for the purposes of the statistical theory, the field $\gr_{\ga}$ as a function only  of $\bfx_0,t$ and let $t_0$ to be taken  at a fixed value. We can also go a step further and note that for every value of $\bfx_0$ the governing equation is an ordinary differential equation. It follows that in constructing an MSR theory for $\gr_\ga$ we have two options: We may  construct a statistical theory for the \emph{restricted problem} in which $\bfx_0$ is also fixed and the field $\gr_\ga$ is taken as a function of only $t$, or a theory for the \emph{full problem} in which only $t_0$ remains fixed and $\gr_\ga$ is taken as a function of $\bfx_0$ and $t$.  In the restricted case we cannot calculate correlations between fields $\gr_{\ga}$ with different values of $\bfx_0$.  In the full case, we can. For our needs, the restricted statistical theory will be sufficient.

 Introduce an operator $\cQ_{\bfx_0}[\gr]$ via the kernel
\begin{equation}
Q_{\ga\gb}^{\bfx_0}[\gr](t,\bfy,\tau) \equiv \gd_{\ga\gb} \gd (t-\tau) \gd (\bfy-\bfx_0-\gr (t)),
\end{equation}
 such that
\begin{equation}
\cQ_{\bfx_0}[\gr] u_{\ga} \equiv \int_{\bbR^d} d\bfy \int_{-\infty}^{+\infty} d\tau \; Q_{\ga\gb}^{\bfx_0}[\gr] (t,\bfy, \tau) u_{\gb} (\bfy, \tau) = u_{\ga}(\bfx_0 + \gr (t), t).
\end{equation}
We also introduce a functional $\cL_{\bfx_0, t_0} [u]$ that constructs $\gr_{\ga}$ from the velocity field. This operation is of course admissible in both the restricted and the full theory.

Since  $\gr = \cL_{\bfx_0, t_0} [u]$ is equivalent to $ \grdot_{\ga} = \cQ_{\bfx_0}[\gr]u_{\ga}$, it follows that there exists a functional $J[u]$ such that $\gd [\grdot_{\ga} - \cQ_{\bfx_0}[\gr]u_{\ga}] = J[u]\gd[\gr - \cL_{\bfx_0, t_0} [u]]$  which can be evaluated by integrating both sides over $\gr$:
  \begin{equation}
J[u] = \int_{\cP (t_0)} \cD\gr\; \gd[\grdot_{\ga} - \cQ_{\bfx_0}[\gr]u_{\ga}]. 
\end{equation}
The integral is a Feynman path integral \cite{book:Hibbs:1965} (a pedagogical introduction is given in Ref. \cite{book:Stevens:1995}). Here, $\cP (t_0)$ is the domain of integration and it is defined as  the set of all $\gr_\ga (t)$ that satisfy the initial condition  $\gr_\ga (t_0) =0$. We also define $\cP$   as the set of all possible paths.   Suppose we would like to evaluate the ensemble average $\langle  M[u,\gr] \rangle$ where $M$ is some arbitrary functional of  $\gr_\ga$ and $u_\ga$. We treat  the velocity field $u_\ga$ as a forcing field with known  statistics.  We assume then that we know how to evaluate the ensemble  average of any expression in terms of the velocity field.  We have:
\begin{align*}
 \langle M[u,\gr] \rangle &= \langle M[u,\cL_{\bfx_0, t_0} [u]]\rangle = \left\langle \int_{\cP (t_0)}\cD\grr\; M[u,\grr]\gd[\grr - \cL_{\bfx_0, t_0} [u]] \right\rangle \\
 &= \int_{\cP (t_0)} \cD\grr\;  \langle M[u,\grr]\gd[\grr - \cL_{\bfx_0, t_0} [u]] \rangle\\
 &= \int_{\cP (t_0)} \cD\grr\;  \langle M[u,\grr]J^{-1}[u] \gd [\grrdot_{\ga} - \cQ_{\bfx_0}[\grr]u_{\ga}] \rangle\\
 &= \int_{\cP (t_0)} \cD\grr\; \left\langle M[u,\grr]J^{-1}[u] \int_{\cP} \cD\gb\; \exp (i\gb_{\ga}(\grrdot_{\ga} - \cQ_{\bfx_0}[\grr]u_{\ga})) \right\rangle\\
 &= \int_{\cP (t_0)} \cD\grr\int_{\cP}\cD\gb\; \exp (i\gb_{\ga}\grrdot_{\ga}) \langle M[u,\grr]J^{-1}[u] \exp (-i\gb_{\ga} \cQ_{\bfx_0}[\grr]u_{\ga})\rangle .
\end{align*}
Here, we have used the convention that repeated Greek indices imply integrating temporal coordinates throughout their  domain in addition to summation of vector components. For example, the expression $\gb_{\ga}\grdot_{\ga}$ is an implicit abbreviation for
\begin{equation}
\gb_{\ga}\grdot_{\ga} = \int_{-\infty}^{+\infty} dt\; \gb_{\ga}(t) \grdot_{\ga}(t).
\end{equation}
We also use the formal representation for the delta functional
\begin{equation}
\gd [u]= \int_{\cP} \cD\gb\; \exp (i\gb_{\ga} u_{\ga}),
\end{equation}
which is valid in the sense of generalized functional distributions.

If the velocity field is incompressible, it can be shown that $J[u]=1$.  A detailed proof of this result is given in appendix \ref{sec:det}. Then, the stochastic theory simplifies to:
 \begin{equation}
 \langle M[u,\gr] \rangle = \iint_{\cP (t_0)\times\cP} \cD\grr\cD\gb\;  \exp  (i\gb_{\ga}\grrdot_{\ga} ) \langle M[u,\grr] \exp (-i\gb_{\ga}  \cQ_{\bfx_0}[\grr]u_{\ga})\rangle . \label{eq:msrtraj}
 \end{equation}
This statement is a concise expression of the statistical theory for
Lagrangian trajectories.

\subsection{Transform back to Eulerian representation}

We now use the statistical theory to derive the relationship between the quasi-Lagrangian correlation and the Eulerian correlation.  The proof given here follows the one given by L'vov and  Procaccia \cite{article:Procaccia:1995:1}, but it is presented in more detail to show the underlying assumptions. The proof  makes an essential use of Eq. \eqref{eq:msrtraj} derived above.

The argument is essentially based on the following identity:
 \begin{equation}
\rho_{\alpha}(\bfr_0 ,t_0 + \Delta t |t) = \gr_\ga (\bfr_0, t_0 | t) - \gr_\ga (\bfr_0, t_0 |t_0 + \gD t).
\label{eq:rhoid}
\end{equation}
To see why this is true, note that the expression on the right-hand side satisfies the governing equation for the $t_0 + \Delta t$ problem, and it also satisfies its initial condition.  Therefore, by uniqueness,  the right-hand side has to be equal to the left-hand side.

To facilitate with calculations, we define $\cM_{\bfx_0}[\gb, \gr]$ as 
\begin{align}
\cM_{\bfx_0}[\gb, \gr] &= \avg{M[u,\gr] \exp (-i\gb_{\ga} \cQ_{\bfx_0}[\gr]u_{\ga})} \\
&= \avg{M[u,\gr] \exp \left(-i\int dt\; \gb_\ga (t) u_\ga (\bfx_0 + \gr (t), t)\right)},
\end{align} 
and we also use the notation $\cM (\bfx_0, t_0)$ for the ensemble average $\avg{M [u,\gr]}$ evaluated under a given choice of $\bfx_0$ and $t_0$.  Consequently, we may write
\begin{align}
\cM (\bfx_0, t_0) &= \iint_{\cP (t_0)\times\cP} \cD\gr \cD\gb \exp (i\gb_\ga \grdot_\ga) \cM_{\bfx_0}[\gb, \gr] \\
&=  \iint_{\cP (t_0)\times\cP} \cD\gr \cD\gb \exp \left(i \int dt\; \gb_\ga (t) \pderiv{\gr_\ga (t)}{t} \right) \cM_{\bfx_0}[\gb, \gr].
\end{align}

The key statement to be proven is the following proposition, that shows the connection between stationarity in the quasi-Lagrangian representation and homogeneity in the Eulerian representation. 

\begin{proposition}
If the velocity field $u$ is incompressible, then
\begin{equation}
\forall \gb \in\cP : \forall \bfx \in \bbR^d : \cM_{\bfx_0}[\gb, \gr + \bfx] = \cM_{\bfx_0}[\gb, \gr] \implies \forall \gD t \in \bbR : \cM (\bfx_0, t_0 + \gD t) = \cM (\bfx_0, t_0) 
\end{equation}
\label{prop:qllemma}
\end{proposition}

\begin{proof}

To facilitate our argument, introduce a new field $\gl_\ga $ defined as equal to the right hand side of \eqref{eq:rhoid}.
\begin{align}
\gl_\ga (\bfr_0, t_0 | t) &= \rho_{\alpha}(\bfr_0 ,t_0 + \Delta t |t) \\ 
&= \gr_\ga (\bfr_0, t_0 | t) - \gr_\ga (\bfr_0, t_0 |t_0 + \gD t) = B_{\ga\gb} \gr_\gb.
\end{align}
 The connection between $\gl_\ga$ and $\gr_\ga$ is \emph{linear}, in the sense that we can construct an appropriate kernel $B_{\ga\gb}$ made of delta functions that transforms one field into the other. The functional  determinant of $B$ is equal to 1, so a change in variables under the path integral does not introduce an additional factor, namely $\cD \gl = \cD\gr$. This is usually true with simple transformations, such as space shifting and rotations, because they merely reshuffle the order in which we integrate over all possible histories. In this case, we need to take into account that the permissible histories are constrained by the initial condition $\gl_\ga (\bfr_0, t_0 |t_0 +\gD t) = 0$ which is different from the initial condition of the field $\gr_\ga$.    It follows that, while there is no need to introduce a functional determinant, the domain of integration has to change from $\cP (t_0)$ to  $\cP (t_0 + \gD t)$ . Finally, it is easy to see that $\pderivin{\gl_\ga}{t} = \pderivin{\gr_\ga}{t}$ and the hypothesis implies that $\cM_{\bfx_0}[\gb, \gr] = \cM_{\bfx_0}[\gb, \gl]$. We may then write:
\begin{align}
\cM (\bfx_0, t_0) &=  \iint_{\cP (t_0)\times\cP} \cD\gr \cD\gb \exp \left(i \int dt\; \gb_\ga (t) \pderiv{\gr_\ga (t)}{t} \right) \cM_{\bfx_0}[\gb, \gr] \\
&=  \iint_{\cP (t_0 + \gD t)\times\cP} \cD\gl \cD\gb \exp \left(i \int dt\; \gb_\ga (t) \pderiv{\gl_\ga (t)}{t} \right) \cM_{\bfx_0}[\gb, \gl]\\  
&= \cM (\bfx_0, t_0 + \gD t) \;\forall \gD t \in \bbR.
\end{align}

\end{proof}

\begin{proposition}
If $u_\ga$ is incompressible, then 
\begin{equation}
\bfu \in\cH^{\ast}_\gw \implies \forall \gD t \in \bbR : W_\ga (\bfx_0, t_0 + \gD t | \bfx, \bfxp, t ) \seqss W_\ga (\bfx_0, t_0 | \bfx, \bfxp, t).
\end{equation}
\label{prop:claim}
\end{proposition}
 
\begin{proof}
Let $n \in \bbN^{\ast}$ be given, and define the functional $M[u,\gr]$ as
\begin{align}
M[u,\gr] &= \prod_{k=1}^n W_{\ga_k}  (\bfx_0, t_0 | \bfx_k, \bfxp_k, t) \\
&= \prod_{k=1}^n [u_{\ga_k} (\bfx_k + \gr (\bfx_0, t_0 | t), t) - u_{\ga_k} (\bfxp_k + \gr (\bfx_0, t_0 | t), t)].
\end{align}
Consequently, the functional $\cM_{\bfx_0} [\gb, \gr]$
reads
\begin{align}
\cM_{\bfx_0} [\gb, \gr] &= \avg{M[u,\gr] \exp\left( -i \int dt \; \gb_\ga (t) u_\ga (\bfx_0 + \gr (t), t) \right)} \\
&= \sum_{m=0}^{+\infty} \frac{(-i)^m}{m!} \idotsint dt_1 \cdots dt_m \;  \gb_{\ga_1} (t_1) \cdots \gb_{\ga_m} (t_m) \notag \\
&\quad \times\avg{\left[  \prod_{l=1}^n W_{\gb_l} (\bfx_0, t_0 | \bfx_l, \bfxp_l, t)\right] \left[ \prod_{k=1}^m u_{\ga_k} (\bfx_0 + \gr (t_k), t_k) \right] }.
\label{eq:crazyavg}
\end{align}
From the assumption $u \in \cH^{\ast}_\gw$ we see that the ensemble average in the equation above is invariant with respect to a uniform spatial shift.  It follows that $\cM_{\bfx_0} [\gb, \gr+ \bfx] = \cM_{\bfx_0} [\gb, \gr], \forall \bfx\in \bbR^d$ , and using proposition \ref{prop:qllemma}, this implies that $\cM (\bfx_0, t_0 + \gD t) = \cM (\bfx_0, t_0), \;\forall \gD t \in \bbR$.  Consequently, we have \begin{align}
\cF_n (\bfx_0, t_0 | \{\bfX\}_n, t) &= \cM (\bfx_0, t_0) = \cM (\bfx_0, t_0+\gD t)\\
&= \cF_n (\bfx_0, t_0 +\gD t | \{\bfX\}_n, t) ,\;\forall n\in \bbNs.
\end{align}
\end{proof}

The claim \eqref{eq:theclaim} follows by combining proposition \ref{prop:claim} with proposition \ref{prop:stat}.  It should be noted that once the relationship between quasi-Lagrangian correlation functions and Eulerian correlation functions is established, it can be easily extended to response functions as well without making any further assumptions.  Starting from the stationarity condition \eqref{eq:eulstat}, we deduce from the quasi-Lagrangian formulation of the Navier-Stokes equations that the quasi-Lagrangian forcing field is also stationary.  Then, we may use an MSR theory on the quasi-Lagrangian Navier-Stokes equations to obtain stationarity on the response functions.  From there, the relationship between the quasi-Lagrangian response functions and the Eulerian response functions can be easily established. 

\subsection{A derivation via conditional local homogeneity}

The artifact introduced by the quasi-Lagrangian formulation is that the turbulent velocity field is being perceived from the viewpoint of an arbitrary fluid particle whose own motion is also stochastic.  Consequently, to relate the quasi-Lagrangian correlation tensor $\cF_n (\bfx_0, t_0 | \{\bfX\}_n, t)$ with the Eulerian correlation tensor $F_n (\{\bfX\}_n, t)$ a certain sense of homogeneity is required to ensure that the velocity field is being perceived by the fluid particle in the same way regardless of the actual position of the particle.  Our analysis of the proof, given previously, has shown that the  homogeneity condition used by the proof is   stronger than the condition $\bfu\in\cH_1$  required to eliminate the sweeping interactions.  What is particularly interesting about the stronger condition $\bfu\in\cH_{\gw}^{\ast}$ is that it requires translational invariance from a group of the \emph{many-time} correlation tensors $F_{m,n}^{\ast}$.

Let us now consider an alternative approach.  Introduce the conditional correlation tensor defined as
\begin{align}
\cF_n (\bfx_0, t_0, \bfy | \{\bfX\}_n, t) &= \avg{\left. \prod_{k=1}^n W_{\ga_k} (\bfx_0, t_0 | \bfx_k, \bfxp_k, t) \right| \rho (\bfx_0, t_0 |t) = \bfy}\\
&= \avg{\left. \prod_{k=1}^n w_{\ga_k} (\bfx_k+\bfy, \bfxp_k+\bfy, t) \right| \rho (\bfx_0, t_0 |t) = \bfy}.
\end{align}
This definition is identical to the definition of the quasi-Lagrangian correlation tensor $\cF_n (\bfx_0, t_0 | \{\bfX\}_n, t)$, except that the ensemble average is replaced with the conditional average predicated on the fluid particle being located at position $\bfy$ at a given time $t$. Let $p (\bfx_0, t_0 | \bfx, t)$ be the probability that a fluid particle originating at $(\bfx_0, t_0)$ will be located at $\bfx$ at time $t$. It follows that the Eulerian correlation tensor $F_n (\{\bfX\}_n, t)$ and the quasi-Lagrangian tensor $\cF_n (\bfx_0, t_0 | \{\bfX\}_n, t)$ are given by
\begin{align}
F_n (\{\bfX\}_n, t) &=  \int d\bfy \; \cF_n (\bfx_0, t_0, \bfy | \{\bfX\}_n - \bfy, t) p (\bfx_0, t_0 | \bfy, t)\\
\cF_n (\bfx_0, t_0 | \{\bfX\}_n, t) &= \int d\bfy \; \cF_n (\bfx_0, t_0, \bfy | \{\bfX\}_n, t) p (\bfx_0, t_0 | \bfy, t).
\end{align}
It is trivial to see that if $\cF_n (\bfx_0, t_0, \bfy | \{\bfX\}_n, t)$ is invariant  with respect to $\{\bfX\}_n \mapsto \{\bfX\}_n+\gD\bfx$ (conditional local homogeneity), then the  Eulerian correlator and the quasi-Lagrangian correlator will be equal.

Now let us consider the implications from a Kolmogorov-like definition of local homogeneity where we assume that $\cF_n (\bfx_0, t_0, \bfy | \{\bfX\}_n, t)$ is independent of $\bfy$ \emph{without} assuming invariance  with respect to $\{\bfX\}_n \mapsto \{\bfX\}_n+\gD\bfx$. Then we have
\begin{align}
\cF_n (\bfx_0, t_0 | \{\bfX\}_n, t) &= \int d\bfy \; \cF_n (\bfx_0, t_0, \bfy | \{\bfX\}_n, t) p (\bfx_0, t_0 | \bfy, t)\\
 &= \cF_n (\bfx_0, t_0, \bfy | \{\bfX\}_n, t) \int d\bfy \; p (\bfx_0, t_0 | \bfy, t)\\
 &= \cF_n (\bfx_0, t_0, \bfy | \{\bfX\}_n, t),
\label{eq:kolmdef2}
\end{align}
and
\begin{align}
F_n (\{\bfX\}_n, t) &=  \int d\bfy \; \cF_n (\bfx_0, t_0, \bfy | \{\bfX\}_n - \bfy, t) p (\bfx_0, t_0 | \bfy, t)\\
 &=  \int d\bfy \; \cF_n (\bfx_0, t_0 | \{\bfX\}_n - \bfy, t) p (\bfx_0, t_0 | \bfy, t).
\label{eq:kolmdef1}
\end{align}
To establish equality between the Eulerian correlation tensor $F_n (\{\bfX\}_n, t)$ and  the quasi-Lagrangian tensor $\cF_n (\bfx_0, t_0 | \{\bfX\}_n, t)$, we also need $\cF_n (\bfx_0, t_0 | \{\bfX\}_n, t)$ to be invariant with respect to $\{\bfX\}_n \mapsto \{\bfX\}_n+\gD\bfx$. Unfortunately, this  can only be established if we also assume an ergodic-like hypothesis that $p (\bfx_0, t_0 | \bfx, t)$ is independent of $\bfx$. It is reasonable to expect this hypothesis to hold for $t \gg t_0$. Then, it follows that 
\begin{align}
\cF_n (\bfx_0, t_0 | \{\bfX\}_n+\gD\bfx, t)  &= \int d\bfy \; \cF_n (\bfx_0, t_0, \bfy | \{\bfX\}_n+\gD\bfx, t) p (\bfx_0, t_0 | \bfy, t)\\
&= \int d\bfy \; \cF_n (\bfx_0, t_0, \bfy +\gD\bfx| \{\bfX\}_n+\gD\bfx, t) p (\bfx_0, t_0 | \bfy+\gD\bfx, t)\\
&=  \int d\bfy \; \cF_n (\bfx_0, t_0, \bfy | \{\bfX\}_n, t) p (\bfx_0, t_0 | \bfy, t) = \cF_n (\bfx_0, t_0 | \{\bfX\}_n, t).
\label{eq:kolmdef3}
\end{align}
This result, combined with Eq. \eqref{eq:kolmdef1} and \eqref{eq:kolmdef2}, implies that
\begin{equation}
\cF_n (\bfx_0, t_0 | \{\bfX\}_n, t)=F_n (\{\bfX\}_n, t),
\label{eq:kolmdef4}
\end{equation}
which in turn, combined with Eq. \eqref{eq:kolmdef3} also gives the $\bfu\in\cH_0$ condition:
\begin{equation}
F_n (\{\bfX\}_n+\gD\bfx, t) = F_n (\{\bfX\}_n, t), \;\forall \gD\bfx\in\bbR^d.
\label{eq:kolmdef5}
\end{equation}
It appears that the ergodic assumption on $ p (\bfx_0, t_0 | \bfx, t)$ is necessary to derive Eqs. \eqref{eq:kolmdef3}, \eqref{eq:kolmdef4}, and  \eqref{eq:kolmdef5}. The assumption of invariance of $\cF_n (\bfx_0, t_0, \bfy | \{\bfX\}_n, t)$ with respect to $\bfy$ is not sufficient.

 As we have mentioned previously, in his first paper, Kolmogorov \cite{article:Kolmogorov:1941} also defined local homogeneity using a conditional ensemble average conditioned on the fluid velocity at the reference point.  Because of the approximate nature of the quasi-Lagrangian transformation used by Kolmogorov (which is not identical to the quasi-Lagrangian transformation of Belinicher and L'vov), his definition can be rephrased in terms of a conditional average on the \emph{location} of the fluid particle. We have shown that  if one uses the Belinicher-L'vov quasi-Lagrangian transformation instead of the Kolmogorov quasi-Lagrangian transformation, and changes the conditional ensemble average from using the \emph{velocity} of the reference fluid particle to using the \emph{location} of the reference fluid particle, then this modified definition of local homogeneity combined with a reasonable ergodic-like hypothesis \emph{does} eliminate the sweeping interactions. We may conjecture that Kolmogorov had the elimination of sweeping  in mind when he formulated his definition, but there is no such explicit indication in his papers.

\section{How the elimination of the sweeping interactions should be justified}

We have seen that when using the quasi-Lagrangian transformation we end up making the homogeneity assumption $\bfu\in\cH_{\gw}^{\ast}$ which is much stronger than the assumption of incremental homogeneity $\bfu\in\cH_0(\cA)$ of the Frisch framework of hypotheses, otherwise we cannot return back to the Eulerian representation.  Furthermore, this homogeneity assumption is introduced implicitly  just by  assuming stationarity in the quasi-Lagrangian representation, even if we don't wish to go back to the Eulerian representation (see proposition \ref{prop:rstat}).  The question that we would like to consider now is whether  the utility of the theoretical work \cite{article:Procaccia:1995:1,article:Procaccia:1995:2,article:Procaccia:1996,article:Procaccia:1997,article:Procaccia:1998,article:Procaccia:1998:1,article:Procaccia:1998:2,article:Procaccia:2000} that relies on the transformation itself is diminished.  We would like to claim that this is not the case, and define a line of investigation that can clarify this further. From an experimental standpoint, the very existence of a robust energy cascade indicates that the sweeping effect is confined to the large scales, and therefore it can be neglected with impunity. The main  question that needs to be addressed is: How should one justify \emph{theoretically} the elimination of the sweeping interactions?

\subsection{Elimination of the sweeping interactions}

 It is widely accepted that the behavior of the structure functions in the inertial range does not depend on the statistical properties of forcing, as long as the spectrum of the forcing term is confined to large length scales.  In a sense, as the energy cascades toward smaller length scales, the characteristic features of the forcing term are ``forgotten''. One may conjecture that the sweeping interactions behave in a similar way as a large-scale forcing term whose effect is forgotten in the inertial range.  We may base this conjecture on the fact that even though the required homogeneity symmetry $\bfu\in\cH_1(\cA)$ may not hold exactly, it can be expected to hold  asymptotically at small scales.  Consequently, even though we cannot set $I_n(\{\bfX\}_n)=0$ exactly, we might expect this term to become rapidly small when the average separation $R$  between the points $\{\bfX\}_n$ goes to zero. But does it vanish rapidly enough? A  rigorous argument would have to estimate how fast $I_n$, as a function of $R$, is approaching zero in the small-scale limit $R/\ell_0\goto 0$, and then calculate the scaling exponent $\gD_n$ associated with the ratio 
\begin{equation}
\frac{I_n (R\{\bfX\}_n)}{(\cO_{nk} F_{n+1})(R\{\bfX\}_n)} \sim \fracp{R}{\ell_0}^{\gD_n}, 
\end{equation}
where $R$ is the scaling parameter and $\ell_0$ the forcing scale. Then, provided that one starts with the assumption $\bfu\in\cH_0$,  proving $\gD_n > 0$ is also a proof that $\bfu\in\cH_1 (\cA)$ which is sufficient to eliminate the sweeping interactions.  Let $\gl_n$ be the scaling exponent of $I_n (R\{\bfX\}_n)$. If we assume that the generalized structure functions $F_n (R\{\bfX\}_n)$ satisfy the fusion rules \cite{article:Procaccia:1996:1,article:Procaccia:1996:3}, then the scaling exponent of  $\cO_{nk} F_{n+1} (R\{\bfX\}_n)$ is $\gz_{n+1}-1$ and it follows that $\gD_n = \gl_n - (\gz_{n+1}-1)$. The challenge, then, is to calculate the scaling exponents $\gl_n$ which are not likely to be universal.

It is easy to see that this argument cannot be extended  to the inverse energy cascade of two-dimensional turbulence.  In that case, the forcing term operates at large wavenumbers. Given that  we can reasonably assume that the inverse energy cascade is local, we expect that the forcing term is forgotten in the inertial range.  The problem is that the energy is now going towards small wavenumbers.  As we have noted,  $I_n$ essentially measures how much homogeneity is violated at a given length scale $R$.  At large length scales, the flow will begin to sense the violation of homogeneity caused by the boundary conditions which will in turn make the sweeping term $I_n$ larger in magnitude. If it becomes comparable to the terms $\cO_{nk} F_{n+1}$, it will probably disrupt the inverse energy cascade.

Numerical simulations have shown that it is possible to obtain an inverse energy cascade under certain conditions \cite{article:Vergassola:2000}, but it can also be disrupted under other conditions \cite{article:Borue:1994,article:Gurarie:2001,article:Gurarie:2001:1,article:Danilov:2003}. Physically, this disruption arises from the spontaneous generation of long-lived coherent vortices that carry a significant amount of enstrophy. An explanation of this effect was given by  Boffetta \emph{et al} \cite{article:Vergassola:2000}, in terms of the ``bottleneck'' effect \cite{article:Falkovich:1994:1}. The general idea is that the behavior of the energy spectrum in the inertial range is modified at wavenumbers near the dissipation range because some of the triad interactions at these length scales are disrupted by the dissipation term, thus making the transfer of energy less efficient.  It is reasonable to anticipate the same effect in a high-resolution simulation of the inverse energy cascade, where the cascade has manifested successfully, without being arrested by coherent structures.  However, we would like to suggest that the deviations observed by Danilov and Gurarie \cite{article:Gurarie:2001,article:Gurarie:2001:1,article:Danilov:2003} and Borue \cite{article:Borue:1994} are more likely to be caused by a similar effect where the triad interactions are disrupted \emph{by the sweeping term $I_n$ rather than the dissipation term $\cD_n F_n$} at large scales. 

The coherent structures that appear  in two-dimensional turbulence can be conceptualized as concentrated small blobs of very high vorticity that raise a two-dimensional  ``hurricane'' in the velocity field around them. Thus, from an intuitive standpoint, it is reasonable to expect that their presence in the flow should amplify the sweeping effect. From a theoretical perspective, one can say that $I_n$ excites a ``particular'' solution of the statistical theory for the correlators $F_n$ which combines linearly \cite{article:Tung:2005,article:Tung:2005:1} with the ``homogeneous'' solution of the homogeneous theory $\cO_{n} F_{n+1}=0$ that corresponds to the inverse energy cascade.  It follows then that to obtain an inverse energy cascade in the forced-dissipative setting, one requires a dissipation term at large scales which will not only dispose of the incoming energy, but will also damp out the sweeping term $I_n$ over the entire range of length scales where it is comparable to $\cO_{nk} F_{n+1}$. It also follows that there should be a conspicuous discrepancy between the energy spectrum in the quasi-Lagrangian representation and the energy spectrum in the Eulerian representation, when the coherent structures provide the dominant contribution to the Eulerian energy spectrum.  It is well-known that in the Eulerian representation the  coherent structures contribute a dominant $k^{-3}$ term \cite{article:Borue:1994}.  On the other hand, in the quasi-Lagrangian representation, one should recover the $k^{-5/3}$ contribution from the underlying inverse energy cascade.  It has already been established that the underlying $k^{-5/3}$ spectrum can be recovered if the coherent structures are artificially removed, either by a wavelet technique \cite{article:Fischer:2005}, or more crudely \cite{article:Borue:1994,article:Gurarie:2001:1}.  If our conjecture holds, then it should be possible to obtain the same effect simply by transforming into the quasi-Lagrangian representation.

We have referenced the inverse energy cascade of two-dimensional turbulence as an example where it is not safe to ``eliminate'' the sweeping interactions.  The criticism of the quasi-Lagrangian formulation by  Mou and Weichman \cite{article:Weichman:1995}  is essentially that it has not been demonstrated that it is ``safe'', in the same sense, to eliminate the sweeping interactions in the downscale energy cascade of three-dimensional turbulence.

\subsection{Alternatives to Lagrangian transformations}
\label{sec:alternatives}

It is possible to use the theoretical work \emph{based} on the quasi-Lagrangian transformation in a way that requires only the assumption $\bfu\in\cH_1(\cA)$ instead of $\bfu\in\cH_\gw^{\ast}$. This can be done via   the following line of argument:  The quasi-Lagrangian formulation modifies the Navier-Stokes equations by redefining the material derivative (see appendix \ref{sec:qlgoveqs}).  The modified equation remains mathematically equivalent to the Navier-Stokes equation because the velocity field is reinterpreted from an Eulerian field into a quasi-Lagrangian field.  It is precisely this reinterpretation which necessitates the stronger assumption $\bfu\in\cH_{\gw}^{\ast}$ to enable a return back to the Eulerian representation.  On the other hand, if we accept the hypothesis that the sweeping interactions can be absorbed into the statistical forcing term, \emph{we can  modify the equation of motion in precisely the same way without changing the interpretation of the velocity field}.  From there, one can derive the same balance equations \eqref{eq:balance} except that one will have $I_n=0$, and consequently the only assumption that is being made implicitly is just $\bfu\in\cH_1(\cA)$. One may then proceed from this Eulerian modified Navier-Stokes equation and develop the  L'vov-Procaccia theory  \cite{article:Procaccia:1995:1,article:Procaccia:1995:2,article:Procaccia:1996,article:Procaccia:1997,article:Procaccia:1998,article:Procaccia:1998:1,article:Procaccia:1998:2,article:Procaccia:2000} with impunity, since the modified governing equation would have the same mathematical form as the quasi-Lagrangian Navier-Stokes equation. 

In geometrical language, the difference between the quasi-Lagrangian transformation and what I propose is the following: In the quasi-Lagrangian transformation we perceive the flow from the point of view of a single fluid particle.  The claim to be established is that the one-time statistical properties of velocity differences should remain invariant when switching between the inertial frame of reference and the non-inertial frame of reference defined by the fluid particle.  My suggestion is to  consider a  transformation where for a given point in space and time $(\bfx, t)$ we perceive the flow from the point of view of whatever fluid particle just happens to be there at $(\bfx_0, t)$.  This leads  to the Navier-Stokes equations for the Eulerian velocity differences $w_{\ga} (\bfx,\bfxp, t)$.  Then, the homogeneity assumption $\bfu\in\cH_1$ is sufficient  to establish the claim  that the one-time statistical properties of velocity differences $w_{\ga} (\bfx,\bfxp, t)$ will remain invariant under a transformation from the inertial frame of reference to a non-inertial frame of reference defined by the fluid particle at $(\bfx_0, t)$.  This claim is in fact mathematically equivalent to the condition $I_n =0$ which follows from $\bfu\in\cH_1$.

In connection with this argument, it is interesting to note that the idea of just modifying the Navier-Stokes equation was considered by Kraichnan \cite{article:Kraichnan:1964} in 1964, who suggested a more crude modification. This modification brute-forces locality in Fourier space by  discarding triad interactions across a wide wavenumber interval and retaining only the local triad interactions. From the same paper we learn that  Kraichnan suspected that there was a relationship between the quasi-Lagrangian transformation of Kolmogorov and the general idea of modifying the Navier-Stokes equation in such a way but noted that bringing that out rigorously is difficult. In my view the quasi-Lagrangian transformation of Belinicher and L'vov, which is different from the Lagrangian transformation used by Kraichnan in his theories, is the key to finding possibly the best way to modify the Navier-Stokes equations in the way that Kraichnan and Kolmogorov intended. 

An alternative argument that was proposed by Yakhot  \cite{article:Yakhot:1981} and used by Giles \cite{article:Giles:2001} to calculate a perturbation expansion for the scaling exponents $\gz_n$ eliminates the sweeping interactions by modifying the statistical theory itself.  This is different from the quasi-Lagrangian formulation and my proposal where the change is made on the governing equation and \emph{then} propagated into the statistical theory.  Again, to justify why one can modify the statistical theory requires the assumption $\bfu\in\cH_1(\cA)$ or an argument justifying the hypothesis that the sweeping interactions can be modeled as large-scale stochastic forcing, which brings us back to the challenge of showing that $\gD_n > 0$.

\subsection{Estimating the scaling exponent $\gD_n$}

The problem of calculating the scaling exponents $\gl_n$ and $\gD_n$ needs to be investigated primarily with numerical simulations and the analysis of experimental data.  However, it is possible to make a speculative theoretical calculation, if we are willing to commit the following  crimes against reality:  First, we  assume that the  velocity field $u_{\ga}(\bfx,t)$ can be modeled as a random gaussian delta-correlated (in time) stochastic field acting at large scales.  Furthermore, we  assume that the  velocity field $u_{\ga}(\bfx,t)$ has an effect on the velocity differences $w_{\ga} (\bfx,\bfxp, t)$ via the sweeping interactions, but completely disregard the reverse effect of the velocity differences on the  velocity field via eddy viscosity, and the fact that  $u_{\ga}(\bfx,t)$ and $w_{\ga} (\bfx,\bfxp, t)$ are obviously constrained by the definition of $w_{\ga} (\bfx,\bfxp, t)$.  In other words, we assume that $w_{\ga} (\bfx,\bfxp, t)$ is advected as a passive scalar by $u_{\ga}(\bfx,t)$ and that $u_{\ga}(\bfx,t)$ can be assumed to be a random gaussian delta-correlated in time  field. Note that $w_{\ga} (\bfx,\bfxp, t)$ is still also forced by $f_\ga$. 

We have shown in Appendix \ref{app:gaussiansweep} that under these assumptions the sweeping term $I_{n} (\{\bfX\}_n, t)$ can be decomposed into three contributions:
\begin{equation}
I_{n} (\{\bfX\}_n, t) = I_{n,(1)} (\{\bfX\}_n, t) + I_{n,(2)} (\{\bfX\}_n, t) + I_{n,(3)} (\{\bfX\}_n, t),
\end{equation}
which are  given by
\begin{align}
I_{n,(1)}^{\ga_1\ga_2\cdots\ga_n} (\{\bfX\}_n, t) &= \avg{\cU_\gb (\{\bfX\}_n, t)}  H_n^{\ga_1\ga_2\cdots\ga_n\gb} (\{\bfX\}_n, t),\\
I_{n,(2)}^{\ga_1\ga_2\cdots\ga_n} (\{\bfX\}_n, t) &=  \sum_{l=1}^n  \sum_{m=1}^n F_{n-1}^{\ga_1\cdots\ga_{m-1}\ga_{m+1}\cdots\ga_n}(\{\bfX\}_n^m, t) I_{\ga_m}(\bfX_m,\bfX_l,t),\\
I_{n,(3)}^{\ga_1\ga_2\cdots\ga_n} (\{\bfX\}_n, t) &= \sum_{l=1}^n \sum_{m=1}^n B_1^{\ga_m\gb} (\bfX_m,\bfX_l,t)  H_{n-1}^{\ga_1\cdots\ga_{m-1}\ga_{m+1}\cdots\ga_n\gb} (\{\bfX\}_n^m).
\end{align}
Here, $H_n^{\ga_1\ga_2\cdots\ga_n}$ , $B_1^{\ga\gb}$ , and $I_{\ga}$ are defined as
\begin{align}
 H_n^{\ga_1\ga_2\cdots\ga_n\gb} (\{\bfX\}_n, t) &= \left[  \sum_{k=1}^n (\pd_{\gb, \bfx_{k}}+\pd_{\gb, \bfxp_{k}} )F_n^{\ga_1\ga_2\cdots\ga_n} (\{\bfX\}_n, t)\right],\\
B_1^{\ga\gb}(\bfX,\bfY,t) &= \avg{\cU_{\gb}(\bfY,t) w_{\ga} (\bfX , t)}-\avg{\cU_{\gb}(\bfY,t))}\avg{w_{\ga} (\bfX , t)},\\
I_{\ga}(\bfX_1,\bfX_2,t)&=\sum_{k=1}^2  (\pd_{\gb, \bfx_{k}}+\pd_{\gb, \bfxp_{k}} ) B_1^{\ga\gb} (\bfX_1,\bfX_2,t).
\end{align}
It is worth noting that the assumption $\bfu\in\cH_0$ implies that $H_n (\{\bfX\}_n, t)=0$ and therefore $I_{n,(1)} (\{\bfX\}_n, t)=I_{n,(3)} (\{\bfX\}_n, t)=0$.  However, we will retain generality and keep all three terms.  It should also be stressed that we are \emph{not} assuming statistical independence between the velocity field $u_{\ga}(\bfx,t)$ and the velocity differences $w_{\ga} (\bfx,\bfxp, t)$.  On the contrary, we assume that the two are related to each other in the sense that the velocity field  $u_{\ga}(\bfx,t)$ is forcing the velocity differences $w_{\ga} (\bfx,\bfxp, t)$ via the sweeping interactions.  However, the case of total statistical independence gives exactly $I_{n} (\{\bfX\}_n, t) = I_{n,(1)} (\{\bfX\}_n, t)$, so it is covered by our argument bellow.

Let $\gl$ be the scaling exponent of $B_1^{\ga\gb}$ such that 
\begin{equation}
B_1^{\ga\gb}(R\bfX,R\bfY,t)\sim (R/\ell_0)^{\gl},
\end{equation} 
for scales in the inertial range. It immediately follows that $I_{\ga}$ scales as 
\begin{equation}
I_{\ga}(R\bfX_1,R\bfX_2,t)\sim (R/\ell_0)^{\gl}(1/R)g(R).
\end{equation} 
 Here, $g(R)$ is a smooth function which represents departure from local homogeneity in the sense $\bfu\in\cH_1$.  Without loss of generality, we associate the scaling exponent $b$ to the function $g(R)$.  The contribution $(1/R)$ arises from the derivatives.  Using a similar line of argument we see that $ H_n$ scales as 
\begin{equation}
 H_n^{\ga_1\ga_2\cdots\ga_n} (R\{\bfX\}_n, t)\sim (R/\ell_0)^{\gz_n}(1/R)f_n(R),
\end{equation}
 where $f_n(R)$ is also a smooth function representing departure from incremental homogeneity  $\bfu\in\cH_0$. We associate the scaling exponent $a_n$ to the function $f_n(R)$. The three contributions to $I_{n} (\{\bfX\}_n, t)$ then scale as 
\begin{align}
I_{n,(1)} (R\{\bfX\}_n, t) &\sim (R/\ell_0)^{\gz_n}(1/R)f_n(R),\\
I_{n,(2)} (R\{\bfX\}_n, t) &\sim (R/\ell_0)^{\gz_{n-1}}(R/\ell_0)^{\gl}(1/R)g(R),\\
I_{n,(3)} (R\{\bfX\}_n, t) &\sim (R/\ell_0)^{\gl}(R/\ell_0)^{\gz_{n-1}}(1/R)f_{n-1}(R),
\end{align}
 and from power counting we find that the corresponding scaling exponents are
\begin{equation}
\gl_{n,1} = \gz_n -1+a_n,\quad
\gl_{n,2} = \gz_{n-1}+\gl -1+b,\quad \text{ and } 
\gl_{n,3} = \gz_{n-1}+\gl -1+a_{n-1}.
\end{equation}
 Using the multifractal formulation, the contribution that supports the \Holder  exponent $h$ gives $\gz_n = nh+\cZ (h)$ , which gives the following evaluation for the scaling exponents $\gD_n$:
\begin{align}
\gD_{n,1}(h) &=  (\gz_n -1+a_n)- (\gz_{n+1}-1) = -h+a_n,\\
\gD_{n,2}(h) &= (\gz_{n-1}+\gl -1+b)-(\gz_{n+1}-1) =-2h+\gl+\gb,\\
\gD_{n,3}(h) &= (\gz_{n-1}+\gl -1+a_{n-1})-(\gz_{n+1}-1) =-2h+\gl+a_{n-1}.
\end{align}

Because the functions $f_n(R)$ and $g(R)$ are smooth, we can Taylor-expand them around $R=0$ and get, to first order, $a_n=b=1$. It is also reasonable to assume that $\gl>0$ since $B_1^{\ga\gb}$ involves a velocity difference. From these evaluations we find that the window for positive  scaling exponents $\gD_n$ is at least $h\in(0,1/2)$. Admittedly, this is a rather narrow interval, even though it is sufficient for the downscale energy cascade of three-dimensional turbulence.  However, the situation is probably a lot better than that.  If we allow negative evaluations of $R$ , which can be defined by reflecting the points $\{\bfX\}_n$ around their center of mass, we may expect that $R=0$ is an extremum and therefore $f_n'(R)=g'(R)=0$. It is easy to show that, using the evaluation $a_n=b=2$, we find that the window for positive scaling exponents $\gD_n$ covers the entire range $h\in(0,1)$ of local scaling exponents.  Although this is somewhat encouraging, the real challenge is to determine what happens in reality and make a comparison of that against the speculative predictions given above.

\section{Discussion and Conclusion}

In the original formulation of his theory, Kolmogorov assumed local homogeneity, local isotropy, and local stationarity in a non-Eulerian representation very similar to the quasi-Lagrangian representation of Belinicher and L'vov. Frisch \cite{article:Frisch:1991,book:Frisch:1995} revised this argument by stating the same assumptions in the Eulerian representation.  This is a decision that we agree with, because the energy spectrum and the structure functions are both Eulerian rather than Lagrangian quantities. Furthermore, as we have argued in the previous section, it is desirable to justify the elimination of the sweeping interactions theoretically, rather than hide the problem under a stronger definition of local homogeneity. Frisch \cite{book:Frisch:1995} has also chosen to strengthen the assumption of self-similarity to make it possible to deduce all the scaling exponents $\gz_n$ and obtain the prediction  $\gz_n=n/3$. Ultimately, this assumption needs to be replaced with a weaker assumption of self-similarity to permit intermittency corrections, and this is the approach followed in the  L'vov-Procaccia theory   \cite{article:Procaccia:1995:1,article:Procaccia:1995:2,article:Procaccia:1996,article:Procaccia:1997,article:Procaccia:1998,article:Procaccia:1998:1,article:Procaccia:1998:2,article:Procaccia:2000}.  Frisch himself proposed the multi-fractal hypothesis \cite{book:Frisch:1995}, which converges with the approach of Belinicher \emph{et al.} in the papers \cite{article:Procaccia:1998,article:Procaccia:1998:1,article:Procaccia:1998:2} in a very interesting way.  Finally, Frisch \cite{book:Frisch:1995} made the very important observation that, in order to carry  Kolmogorov's argument through, it is necessary to assume the existence of an anomalous energy sink.

In the present paper we have shown that  the assumptions of the Frisch framework are still not strong enough to prove the $4/5$-law in the Eulerian representation. We have also shown that the problem of eliminating the sweeping interactions with a predictive argument remains open. The hypothesis $\bfu\in\cH_1 (\cA)$ can rectify both problems. Even better, starting from the more reasonable hypothesis $\bfu\in\cH_0(\cA)$, a rigorous proof that establishes $\gD_n > 0$ would be sufficient to establish $\bfu\in\cH_1 (\cA)$.  A positive response to the question raised recently by Frisch \cite{article:Frisch:2005} concerning the self-consistency of incremental homogeneity in the $\bfu\in\cH_0(\cA)$ sense  would also follow from the validity of the conjecture $\gD_n > 0$. This would be a fundamental breakthrough finally putting to rest the problem of the sweeping interactions that has concerned the community for the last 60 years.  It would essentially establish that  the sweeping interactions can be modeled as stochastic forcing acting only at large scales.  Then we can simply drop from the Navier-Stokes equations the portion of the nonlinearity associated with the sweeping interactions, and build the entire statistical theory on the modified Navier-Stokes equations. We have also explained why it may not be desirable to use  the Belinicher-L'vov quasi-Lagrangian formulation to go around the problem.  The reason is that using the quasi-Lagrangian formulation requires the even stronger assumption $\bfu\in\cH_{\gw}^{\ast}$.

We would also like to emphasize that our conclusions are \emph{not} a criticism of the  L'vov-Procaccia theory \emph{based} on the quasi-Lagrangian formulation \cite{article:Procaccia:1995:1,article:Procaccia:1995:2,article:Procaccia:1996,article:Procaccia:1997,article:Procaccia:1998,article:Procaccia:1998:1,article:Procaccia:1998:2,article:Procaccia:2000}. In fact, as long as one's intention is to solve the problem of  \emph{globally} homogeneous  turbulence, there is no issue whatsoever with respect to the sweeping interactions  and the quasi-Lagrangian transformation, provided that the assumption of global homogeneity is \emph{many-time} rather than \emph{one-time}.  On the other hand, it is desirable to move away from the assumptions of global homogeneity and global isotropy, which cannot be physically realized, and take steps towards building a theory based on the assumptions of asymptotic incremental homogeneity $\bfu\in\cH_0(\cA)$ and incremental isotropy  in an Eulerian framework, as envisioned by Frisch.  Our paper implies that the results of the  L'vov-Procaccia theory  \cite{article:Procaccia:1995:1,article:Procaccia:1995:2,article:Procaccia:1996,article:Procaccia:1997,article:Procaccia:1998,article:Procaccia:1998:1,article:Procaccia:1998:2,article:Procaccia:2000} can be readily carried over and applied towards this goal, provided that the hypothesis $\gD_n > 0$ is proved, and our proposal of section \ref{sec:alternatives} rather than the quasi-Lagrangian transformation is  employed.

\section*{Acknowledgements}

It is a pleasure to thank Ka-Kit Tung for his advice and encouragement and continued interest in this work. The paper was  considerably improved by comments and discussions with Reginald Hill and helpful remarks from an anonymous referee.  This research has been supported in part by the National Science Foundation under the grant  DMS 03-27658.

\appendix

\section{Quasi-Lagrangian representation of the Navier-Stokes equations}
\label{sec:qlgoveqs} 

We show how the quasi-Lagrangian transformation makes it possible to write the Navier-Stokes equations in terms of velocity differences, thereby eliminating the sweeping interactions.  The reader should compare this argument with the derivation of the $4/5$-law by Monin and Yaglom \cite{book:Yaglom:1975}, to see that they are doing the same thing.

We begin by noting that the Eulerian velocity field $u_\ga (\bfx, t)$ can be reconstructed from $v_\ga (\bfx_0, t_0 | \bfx, t)$ as:
\begin{align}
u_\ga (\bfx, t) &= u_\ga (\bfx - \gr (\bfx_0, t_0 | t) +  \gr (\bfx_0, t_0 | t), t) = v_\ga (\bfx_0, t_0 | \bfx -  \gr (\bfx_0, t_0 | t), t).
\end{align}
To eliminate $\gr (\bfx_0, t_0 | t)$ we use
\begin{align}
 \gr_\ga (\bfx_0, t_0 | t) &= \int_{t_0}^t d\tau \; u_\ga (\bfx + \gr (\bfx_0, t_0 | \tau), \tau) = \int_{t_0}^t d\tau \; v_\ga (\bfx_0, t_0 | \bfx, \tau),
\end{align}
and therefore, $u_\ga (\bfx, t)$ reads
\begin{equation}
u_\ga (\bfx, t) = v_\ga \left( \bfx_0, t_0 | \bfx - \int_{t_0}^t d\tau \; \bfv (\bfx_0, t_0 | \bfx, \tau), t \right).
\end{equation}
Let $R_\ga (\bfx_0, t_0 | \bfx, t) $ be defined as
\begin{equation}
R_\ga (\bfx_0, t_0 | \bfx, t) = (\bfx)_\ga - \int_{t_0}^t d\tau \; v_\ga (\bfx_0, t_0 | \bfx, \tau) = (\bfx)_\ga - \gr_\ga (\bfx_0, t_0 | t),
\end{equation}
such that $u_\ga (\bfx, t) = v_\ga (\bfx_0, t_0 | \bfR  (\bfx_0, t_0 | \bfx, t), t)$. Also, define $v_\ga^0  (\bfx_0, t_0 | t) \equiv v_\ga  (\bfx_0, t_0 | \bfx_0, t)$ such that we may write concisely $w_\ga = v_\ga -  v_\ga^0$. It is easy to see that $\pd_{\ga, \bfx} R_\gb = \gd_{\ga\gb}$ and $\pderivin{R_\ga}{t} = -v_\ga^0$, and we may use these relations to show that the quasi-Lagrangian transformation preserves incompressibility, as follows:
\begin{align}
\pd_{\ga,\bfx} w_\ga &= \pd_{\ga,\bfx} (v_\ga - v_\ga^0) = \pd_{\ga,\bfx} v_\ga = (\pd_{\gb,\bfx} v_\ga)\gd_{\ga\gb} = (\pd_{\gb,\bfx} v_\ga) (\pd_{\ga,\bfx} R_\gb) = \pd_{\ga,\bfx} u_\ga =0.
\end{align}

The key result is that the sweeping interactions are eliminated in the transformation of the material derivative itself.  The show this, consider an arbitrary field $U (\bfx, t)$ and its quasi-Lagrangian transformation $\cU (\bfx_0, t_0 | \bfx, t)$ (where the fluid particle follows the Eulerian velocity field $u_\ga (\bfx, t)$ ). From the relation
\begin{equation}
U (\bfx, t) = \cU (\bfx_0, t_0 | \bfR  (\bfx_0, t_0 | \bfx, t), t),
\end{equation}
we find that 
\begin{equation}
\pderiv{U}{t} = \pderiv{\cU}{t} + (\pd_{\ga,\bfx}\cU)\pderiv{R_\ga}{t} = \pderiv{\cU}{t} + (\pd_{\ga,\bfx}\cU)(-v_\ga^0),
\end{equation}
and 
\begin{equation}
\pd_{\ga,\bfx} U  = (\pd_{\gb,\bfx}     \cU)(\pd_{\ga,\bfx} R_\gb) =  (\pd_{\gb,\bfx} \cU) \gd_{\ga\gb} = (\pd_{\ga,\bfx} \cU),
\end{equation}
and it follows that:
\begin{equation}
\Dderiv{U} &= \pderiv{U}{t} + u_\ga\pd_{\ga,\bfx} U = \pderiv{\cU}{t} + w_\ga\pd_{\ga,\bfx} \cU.
\end{equation}
This equation is identical to the unnumbered equation preceding equation (22.14) in Monin and Yaglom \cite{book:Yaglom:1975}.  It is easy to see that since the material derivative is written in terms of velocity differences, if it is applied on  $w_\ga (\bfx_0, t_0 | \bfx, t)$ we shall obtain an equation written exclusively in terms of velocity differences.

\section{Evaluation of $J[u]$}
\label{sec:det}

In this appendix, we provide a detailed evaluation of the functional determinant $J[u]$ that we encounter in the derivation of the MSR theory for Lagrangian trajectories. The procedure was outlined briefly in L'vov and  Procaccia \cite{article:Procaccia:1995:1}. However it is not as trivial as it seems.  We have followed the outline and rederived the following more complete version of the proof:

  First, we discretize time in $\gD t$ intervals and introduce the following notation:
 \begin{equation}
 \begin{split}
 t_n &= t_0 + n\gD t \\ \gr_{\ga}^n &= \gr_{\ga} (\bfx_0, t_0 | t_n) \\
 u_{\ga}^n &= u_{\ga} (\bfx_0 + \gr_{\ga}^n, t_n) \\ u_{\ga}^{n,m} &=
 u_{\ga} (\bfx_0 + \gr_{\ga}^n, t_m).
 \end{split}
 \end{equation}
 Each of the objects $\gr_\ga^n, u_\ga^n, u_\ga^{n,m}$ is a field that is a function of $\bfx_0$ only.
 Note that $\gr_\ga^0 =0$. The governing equation for the Lagrangian trajectories field is equivalent to a
 set of the following discretized equations:
 \begin{equation}
 \frac{\gr_{\ga}^{n+1} - \gr_{\ga}^n}{\gD t} = u_{\ga}^n.
 \end{equation}
There are, of course, many alternative discretizations to choose from.
The rule  is that, once we have chosen a discretization, we
have to stay with it.  We cannot switch to another scheme in the
middle of the computations, for the sake of convenience.  To evaluate
$J[u]$ we proceed from the path integral definition:
\begin{equation}
 \begin{split}
 J[u] &= \int_{\cP (t_0)} \cD\gr\; \gd [\grdot_{\ga} - Q_{\ga\gb}^{\bfx_0}[\gr]u_{\gb}]\\ &=
\lim_{\gD t \rightarrow 0} \prod_{n \in \Bbb{Z}-\{ 0\}}\int \frac{d\gr^n}{a} \gd
 \left(\frac{\gr_{\ga}^{n+1} - \gr_{\ga}^n}{\gD t} - u_{\ga}^n
 \right)\\ &=\lim_{\gD t \rightarrow 0} \prod_{n=1}^{+\infty} \frac{A_n(\gD t)}{a}
 \prod_{n=-\infty}^{-1} \frac{B_n(\gD t)}{a},
\label{eq:Jprod}
 \end{split}
 \end{equation}
where $A_n (\gD t)$ and $B_n (\gD t)$ are defined as
 \begin{equation}
 \begin{split}
 A_n(\gD t) &= \int d\gr_{\ga}^{n+1}\; \gd \left(\frac{\gr_{\ga}^{n+1} -
 \gr_{\ga}^n}{\gD t} - u_{\ga}^n \right), \\ B_n(\gD t) &= \int d\gr_{\ga}^n\;
 \gd \left(\frac{\gr_{\ga}^{n+1} - \gr_{\ga}^n}{\gD t} - u_{\ga}^n
 \right).
 \end{split}
 \end{equation}
Here, $a$ is a normalization constant such that the product in \eqref{eq:Jprod} 
 converges. Obviously, if such a constant exists, it will be unique.

  The $A_n$ integral is easy to evaluate:
 \begin{equation}
 A_n(\gD t) = \int d\gr^{n+1}\; \gD t \gd (\gr_{\ga}^{n+1} - \gr_{\ga}^n -
 \gD t u_{\ga}^n)= \gD t.
 \end{equation}
 To evaluate $B_n$ we need to rewrite the discretized governing
 equation so that it is explicit with respect to $\gr_\ga^n$:
 \begin{equation}
 \begin{split}
 \gr_{\ga}^{n+1} - \gr_{\ga}^n &= \gD t u_{\ga}(\bfx_0 + \gr_{\ga}^n,
 t_n) \\ &= \gD t u_{\ga}(\bfx_0 + \gr_{\ga}^{n+1} - \grdot_{\ga}^{n+1}
 \gD t + O(\gD t^2), t_n) \\ &= \gD t(u_{\ga}(\bfx_0 + \gr_{\ga}^{n+1},
 t_n) - \grdot_{\gb}^{n+1} \gD t \partial_{\gb} (u_{\ga}(\bfx_0 +
 \gr_{\ga}^{n+1}, t_n)) + O(\gD t^2)\\
 &= u_{\ga}^{n+1, n} \gD t - (\gr_{\gb}^{n+1} - \gr_{\gb}^n)
 \partial_{\gb} u_{\ga}^{n+1, n} \gD t + O(\gD t^2),
 \end{split}
 \end{equation}
 therefore
 \begin{equation}
 (\gd_{\ga\gb}+ \partial_{\gb} u_{\ga}^{n+1, n} \gD t)
 (\gr_{\gb}^{n+1} - \gr_{\gb}^n) = u_{\ga}^{n+1, n} \gD t.
 \end{equation}
 We proceed by employing the following change of variables:
  \begin{equation}
 R_{\ga}^n = (\gd_{\ga\gb} + \partial_{\gb} u_{\ga}^{n+1, n} \gD t)
 \gr_{\gb}^n.
 \end{equation}
 The  integral differentials are transformed according to a
  determinant as follows:
 \begin{equation}
dR_{\ga}^n = \det (\gd_{\ga\gb} + \partial_{\gb} u_{\ga}^{n+1, n}
 \gD t)  d\gr_{\gb}^n.
 \end{equation}
 We will now show that incompressibility implies that the 
 determinant is equal to 1.
 For brevity, introduce
 \begin{equation}
 M_{\ga\gb} = \gd_{\ga\gb} + \partial_{\gb} u_{\ga}^{n+1, n}  \gD
 t.
 \end{equation}
   In the determinant expansion, every term other than $M_{11} M_{22}
   M_{33}$ is $O (\gD t^2)$ because it includes at least two off-diagonal
 factors each of which contributes a factor of $\gD t$ . It follows that:

\begin{equation}
 \begin{split}
 \det M &= M_{11} M_{22} M_{33} + O (\gD t^2) \\ &= (1 + \gD t
 \partial_1 u_1^{n+1, n} ) (1 + \gD t \partial_2 u_2^{n+1, n}
) (1 + \gD t \partial_3 u_3^{n+1, n}) + O (\gD t^2) \\ &=
 1 + \gD t (\partial_1 u_1^{n+1, n} + \partial_2 u_2^{n+1, n}
 + \partial_3 u_3^{n+1, n}) + O (\gD t^2)\\ &= 1 + O (\gD
 t^2).
 \end{split}
 \end{equation}

 Note that in the last step we employed the incompressibility condition.
 It follows that  $d \gr^n = d R^n$ .  We may now proceed and evaluate the
 integral $B_n$.
 \begin{equation}
 \begin{split}
 B_n(\gD t) &= \int d\gr^n\; \gD t \gd[\gr_{\ga}^{n+1} - \gr_{\ga}^n - \gD t u_{\ga}^n] \\
 &= \gD t \int d\gr^n\; \gd((\gd_{\ga\gb} + \partial_{\gb} u_{\ga}^{n+1, n} \gD t) (\gr_{\gb}^{n+1} - \gr_{\gb}^n) - u_{\ga}^{n+1, n} \gD t)\\
 &= \gD t (1+O(\gD t^2))\\ &\quad\times \int dR^n\; \gd((\gd_{\ga\gb} + \partial_{\gb} u_{\ga}^{n+1, n} \gD t)\gr_{\gb}^{n+1} - R_{\ga}^n - \gD t u_{\ga}^{n+1, n}) \\
 &= \gD t + O(\gD t^3).
 \end{split}
 \end{equation}
In the last step, the crucial requirement is that $\gr_\gb^{n+1}$ and $u_\ga^{n+1, n}$ should  not depend   on
$\gr_{\ga}^n$ and therefore $R_{\ga}^n$.  If we set the normalization
constant $a=\gD t$, then $J[u]$ evaluates as:
 \begin{equation}
  J[u]= \lim_{\gD t \rightarrow 0}\prod_{n=1}^{+\infty} \frac{A_n(\gD t)}{a} \prod_{n=-\infty}^{-1}
 \frac{B_n(\gD t)}{a} = 1.
 \end{equation}
Note that $O (\gD t^2)$ contributions to the integrals $A_n(\gD t)$ and $B_n(\gD t)$, which we have disregarded, would vanish anyway after taking the limit $\gD t \rightarrow 0$, so they can be safely ignored with impunity.

\section{Sweeping interactions under a gaussian mean field}
\label{app:gaussiansweep}

We exploit the following mathematical result: if $f_{\ga}(\bfx_1,t_1)$ is a Gaussian stochastic field, the ensemble averages of the form $\avg{f_{\ga}(\bfx_1,t_1)R[f]}$ can be evaluated for any analytic functional  $R[f]$  by the following integral
\begin{equation}
\avg{f_{\ga}(\bfx_1,t_1)R[f]} = \avg{f_{\ga}(\bfx_1,t_1)}\avg{R[f]} +\int d\bfx_2 dt_2 \;\avg{f_{\ga}(\bfx_1, t_1)f_{\gb}(\bfx_2, t_2)}_c \avg{\frac{\gd R[f]}{\gd f_{\gb}(\bfx_2, t_2)}},
\label{eq:Novikov}
\end{equation}
where
\begin{equation}
\avg{f_{\ga}(\bfx_1, t_1)f_{\gb}(\bfx_2, t_2)}_c \equiv \avg{f_{\ga}(\bfx_1, t_1)f_{\gb}(\bfx_2, t_2)}-\avg{f_{\ga}(\bfx_1, t_1)}\avg{f_{\gb}(\bfx_2, t_2)}.
\end{equation}
This a generalization of Gaussian integration by parts, a technique attributed by Frisch \cite{book:Frisch:1995} to Novikov \cite{article:Novikov:1965}, Donsker \cite{article:Donsker:1964}  and Furutsu \cite{article:Furutsu:1963}.

We begin the proof by defining the following correlation functions:
\begin{align}
U_{\ga\gb} (\bfx_1, t_1; \bfx_2, t_2) &= \avg{u_{\ga}(\bfx_1, t_1)u_{\gb}(\bfx_2, t_2)}-\avg{u_{\ga}(\bfx_1, t_1)}\avg{u_{\gb}(\bfx_2, t_2)}, \\
B_n^{\ga_1\cdots\ga_n\gb}(\{\bfX\}_n,\bfY,t) &= \avg{(\cU_{\gb}(\bfY,t))\left[  \prod_{l=1}^n w_{\ga_l} (\bfX_{l} , t) \right]}-\avg{\cU_{\gb}(\bfY,t))}F_n^{\ga_1\ga_2\cdots\ga_n} (\{\bfX\}_n,t),\\
H_n^{\ga_1\ga_2\cdots\ga_n\gb} (\{\bfX\}_n, t) &= \left[  \sum_{k=1}^n (\pd_{\gb, \bfx_{k}}+\pd_{\gb, \bfxp_{k}} )F_n^{\ga_1\ga_2\cdots\ga_n} (\{\bfX\}_n, t)\right],
\end{align}
and also the following  response functions:
\begin{align}
R_{\ga\gb} &= \avg{\frac{\gd w_\ga (\bfX, t_1)}{\gd u_\gb (\bfy, t_2)}},\\
R_n^{\ga_1\cdots\ga_n\gb}(\{\bfX\}_n,t,\bfy,\tau) &= \avg{\frac{\gd}{\gd u_\gb (\bfy,\tau)}\left[ \prod_{l=1}^n w_{\ga_l} (\bfX_l,t) \right]}.
\end{align}
Here, we disregard the fact that $u_{\ga}(\bfx,t)$ and $w_{\ga} (\bfx,\bfxp, t)$ are related by definition and assume that the only effect of $u_{\ga}(\bfx,t)$ on $w_{\ga} (\bfx,\bfxp, t)$ is via the sweeping interaction.  We also assume that the velocity field $u_{\ga}(\bfx,t)$ is delta-correlated which implies that
\begin{equation}
U_{\ga\gb} (\bfx_1, t_1; \bfx_2, t_2) = U_{\ga\gb} (\bfx_1,\bfx_2) \gd (t_1-t_2).
\end{equation}
We begin by splitting $I_{n} (\{\bfX\}_n, t)$ into two terms 
\begin{equation}
I_{n} (\{\bfX\}_n, t) = I_{n,(1)} (\{\bfX\}_n, t) + I_{n,(2+3)} (\{\bfX\}_n, t),
\end{equation}
with $I_{n,(1)} (\{\bfX\}_n, t)$ given by
\begin{align}
I_{n,(1)}^{\ga_1\ga_2\cdots\ga_n} (\{\bfX\}_n, t) &= \sum_{k=1}^n (\pd_{\gb, \bfx_{k}}+\pd_{\gb, \bfxp_{k}}) \left\{ \avg{\cU_\gb (\{\bfX\}_n, t)}\avg{\left[  \prod_{l=1}^n w_{\ga_l} (\bfX_{l} , t) \right]}\right\}\\
&=\avg{\cU_\gb (\{\bfX\}_n, t)} \left[  \sum_{k=1}^n (\pd_{\gb, \bfx_{k}}+\pd_{\gb, \bfxp_{k}} )F_n^{\ga_1\ga_2\cdots\ga_n} (\{\bfX\}_n, t)\right]\\
&=\avg{\cU_\gb (\{\bfX\}_n, t)} H_n^{\ga_1\ga_2\cdots\ga_n\gb} (\{\bfX\}_n, t).
\end{align}
Here we have used the  incompressibility condition.
\begin{equation}
\sum_{k=1}^n (\pd_{\gb, \bfx_{k}}+\pd_{\gb, \bfxp_{k}}) \avg{\cU_\gb (\{\bfX\}_n, t)} =0.
\end{equation}
The remaining contribution to $I_{n} (\{\bfX\}_n, t)$ reads
\begin{equation}
I_{n,(2+3)}^{\ga_1\ga_2\cdots\ga_n} (\{\bfX\}_n, t) = \frac{1}{2n}\sum_{k=1}^n \sum_{l=1}^n  (\pd_{\gb, \bfx_{k}}+\pd_{\gb, \bfxp_{k}})  B_n^{\ga_1\cdots\ga_n\gb}(\{\bfX\}_n,\bfX_l,t).
\end{equation}
Using Gaussian integration by parts we may write
\begin{align}
B_1^{\ga\gb} (\bfX,\bfY,t) &= \int d\bfz d\tau\; R_{\ga\gc} (\bfX, t; \bfz,\tau) [U_{\gb\gc} (\bfy,t; \bfz, \tau) +U_{\gb\gc} (\bfyp,t; \bfz, \tau)] \\
&= \int d\bfz \; R_{\ga\gc} (\bfX, t; \bfz, t) [U_{\gb\gc} (\bfy,\bfz) +U_{\gb\gc} (\bfyp,\bfz)] ,
\end{align}
and
\begin{align}
B_n^{\ga_1\cdots\ga_n\gb} (\{\bfX\}_n,\bfY,t) &= \int d\bfz d\tau\; R_n^{\ga_1\cdots\ga_n\gc} (\{\bfX\}_n, t; \bfz,\tau) [U_{\gb\gc} (\bfy,t; \bfz, \tau) +U_{\gb\gc} (\bfyp,t; \bfz, \tau)] \\
&= \int d\bfz \; R_n^{\ga_1\cdots\ga_n\gc} (\{\bfX\}_n, t; \bfz,t)  [U_{\gb\gc} (\bfy,\bfz) +U_{\gb\gc} (\bfyp,\bfz)].
\end{align}
The key step is to note that
\begin{align}
R_{n}^{\ga_1\cdots\ga_n\gb}(\{\bfX\}_n,t,\bfy,t) &= \avg{\frac{\gd}{\gd u_\gb (\bfy,t)}\left[ \prod_{l=1}^n w_{\ga_l} (\bfX_l,t) \right]} \\
&= \sum_{k=1}^n \avg{\left[ \prod_{l=1, l\neq k}^n w_{\ga_l} (\bfX_l,t) \right]\frac{\gd w_{\ga_k}(\bfX_k,t)}{\gd u_\gb (\bfy,t)}}\\
&= \sum_{k=1}^n F_{n-1}^{\ga_1\cdots\ga_{k-1}\ga_{k+1}\cdots\ga_n} (\{\bfX\}_n^k) R_{\ga_k\gb}(\bfX_k, t;\bfy,t).
\end{align}
Here we exploit the fact, first pointed out in Ref. \cite{article:Procaccia:1996}, that the variational derivative $(\gd w_{\ga_k}(\bfX_k,t))/(\gd u_\gb (\bfy,t))$ is not correlated with the velocity differences $w_{\ga_l} (\bfX_l,t)$ because no time is being allowed for the interaction to develop a correlation. This relationship between the response functions  implies a corresponding relationship between $B_n (\{\bfX\}_n,\bfY,t)$ and $B_1 (\bfX_k,\bfY,t)$:
 \begin{align}
 B_n^{\ga_1\cdots\ga_n\gb} (\{\bfX\}_n,\bfY,t) &=\int d\bfz \; \left[ \sum_{k=1}^n F_{n-1}^{\ga_1\cdots\ga_{k-1}\ga_{k+1}\cdots\ga_n} (\{\bfX\}_n^k) R_{\ga_k\gb}(\bfX_k, t;\bfz,t) \right]  [U_{\gb\gc} (\bfy,\bfz) +U_{\gb\gc} (\bfyp,\bfz)] \\
 &=\sum_{k=1}^n F_{n-1}^{\ga_1\cdots\ga_{k-1}\ga_{k+1}\cdots\ga_n} (\{\bfX\}_n^k) \left[ \int d\bfz \;R_{\ga_k\gb}(\bfX_k, t;\bfz,t)[U_{\gb\gc} (\bfy,\bfz) +U_{\gb\gc} (\bfyp,\bfz)]\right]\\
 &=\sum_{k=1}^n F_{n-1}^{\ga_1\cdots\ga_{k-1}\ga_{k+1}\cdots\ga_n} (\{\bfX\}_n^k) B_1^{\ga_k\gb} (\bfX_k,\bfY,t).
 \end{align}
It immediately follows that
\begin{equation}
I_{n,(2+3)}^{\ga_1\ga_2\cdots\ga_n} (\{\bfX\}_n, t) = \frac{1}{2n}\sum_{k=1}^n \sum_{l=1}^n \sum_{m=1}^n (\pd_{\gb, \bfx_{k}}+\pd_{\gb, \bfxp_{k}}) F_{n-1}^{\ga_1\cdots\ga_{m-1}\ga_{m+1}\cdots\ga_n} (\{\bfX\}_n^m) B_1^{\ga_m\gb} (\bfX_m,\bfX_l,t),
\end{equation}
which can be broken down to
\begin{align}
I_{n,(2)}^{\ga_1\ga_2\cdots\ga_n} (\{\bfX\}_n, t) &= \sum_{l=1}^n \sum_{m=1}^n F_{n-1}^{\ga_1\cdots\ga_{m-1}\ga_{m+1}\cdots\ga_n} (\{\bfX\}_n^m) I_{\ga_m} (\bfX_m,\bfX_l,t),\\
I_{n,(3)}^{\ga_1\ga_2\cdots\ga_n} (\{\bfX\}_n, t) &= \sum_{l=1}^n \sum_{m=1}^n B_1^{\ga_m\gb} (\bfX_m,\bfX_l,t) H_{n-1}^{\ga_1\cdots\ga_{m-1}\ga_{m+1}\cdots\ga_n\gb} (\{\bfX\}_n^m),
\end{align}
with
\begin{equation}
I_{\ga}(\bfX_1,\bfX_2,t)=\sum_{k=1}^2  (\pd_{\gb, \bfx_{k}}+\pd_{\gb, \bfxp_{k}} ) B_1^{\ga\gb} (\bfX_1,\bfX_2,t).
\end{equation}

\bibliography{references,references-submit}

\begin{thebibliography}{100}
\expandafter\ifx\csname url\endcsname\relax
  \def\url#1{\texttt{#1}}\fi
\expandafter\ifx\csname urlprefix\endcsname\relax\def\urlprefix{URL }\fi

\bibitem{book:Richardson:1922}
L.~Richardson, Weather prediction by numerical process, Cambridge University
  Press, Cambridge, 1922.

\bibitem{article:Kolmogorov:1941}
A.~Kolmogorov, The local structure of turbulence in incompressible viscous
  fluid for very large {R}eynolds numbers, Dokl. Akad. Nauk. SSSR 30 (1941)
  301--305, english translation published in volume 434 of \textsl{Proc. R.
  Soc. Lond. A}.

\bibitem{article:Kolmogorov:1941:1}
A.~Kolmogorov, Dissipation of energy in the locally isotropic turbulence, Dokl.
  Akad. Nauk. SSSR 32 (1941) 16--18, english translation published in volume
  434 of \textsl{Proc. R. Soc. Lond. A}.

\bibitem{article:Batchelor:1947}
G.~Batchelor, Kolmogorov's theory of locally isotropic turbulence, Proc. Camb.
  Phil. Soc. 43 (1947) 533--559.

\bibitem{article:A.Moilliet:1962}
H.~Grant, R.~Stewart, A.Moilliet, Turbulence spectra from a tidal channel, J.
  Fluid. Mech. 12 (1962) 241--263.

\bibitem{article:Gibson:1962}
M.~Gibson, Spectra of turbulence in a round jet, J. Fluid. Mech. 15 (1962)
  161--173.

\bibitem{book:Frisch:1995}
U.~Frisch, Turbulence: The legacy of {A.N. K}olmogorov, Cambridge University
  Press, Cambridge, 1995.

\bibitem{article:Antonia:1997}
K.~Sreenivasan, R.~Antonia, The phenomenology of small-scale turbulence, Ann.
  Rev. Fluid Mech. 29 (1997) 435--472.

\bibitem{article:Kolmogorov:1962}
A.~Kolmogorov, A refinement of previous hypotheses concerning the local
  structure of turbulence in a viscous incompressible fluid at high {R}eynolds
  number, J. Fluid. Mech. 13 (1962) 237.

\bibitem{article:Oboukhov:1962}
A.~Oboukhov, Some specific features of the atmospheric turbulence, J. Fluid.
  Mech. 13 (1962) 77--81.

\bibitem{lect:Procaccia:1997}
V.~L'vov, I.~Procaccia, Hydrodynamic turbulence: a 19th century problem with a
  challenge for the 21st century, in: O.~Boratav, A.~Eden, A.~Erzan (Eds.),
  Turbulence modeling and vortex dynamics, Proceedings of a workshop held at
  {I}nstabul, Turkey, Springer-Verlag, Berlin, 1997.

\bibitem{article:Frisch:2005}
U.~Frisch, J.~Bec, E.~Aurell, Locally homogeneous turbulence: Is it a
  consistent framework?, Phys. Fluids 17 (2005) 081706.

\bibitem{article:Procaccia:1995:1}
V.~L'vov, I.~Procaccia, Exact resummations in the theory of hydrodynamic
  turbulence. {I}. {T}he ball of locality and normal scaling, Phys. Rev. E 52
  (1995) 3840--3857.

\bibitem{article:Nelkin:1994}
M.~Nelkin, Universality and scaling in fully developed turbulence, Adv. Phys.
  43 (1994) 143--181.

\bibitem{article:Sreenivasan:1999}
K.~Sreenivasan, Fluid turbulence, Rev. Mod. Phys. 71 (1999) S383--S395.

\bibitem{article:Taylor:1935}
G.~Taylor, Statistical theory of turbulence. {P}arts 1-4, Proc. Roy. Soc. A 151
  (1935) 421--478.

\bibitem{article:Taylor:1936}
G.~Taylor, Statistical theory of turbulence. {P}art 5, Proc. Roy. Soc. A 156
  (1936) 307--317.

\bibitem{article:Howarth:1938}
T.~Karman, L.~Howarth, On the statistical theory of isotropic turbulence, Proc.
  Roy. Soc. A 164 (1938) 192--215.

\bibitem{article:Robertson:1940}
H.~Robertson, The invariant theory of isotropic turbulence, Proc. Camb. Phil.
  Soc. 36 (1940) 209--223.

\bibitem{book:Batchelor:1953}
G.~Batchelor, The theory of Homogeneous Turbulence, Cambridge University Press,
  Cambridge, 1953.

\bibitem{book:Leslie:1972}
D.~Leslie, Developments in the theory of turbulence, Clarendon Press, Oxford,
  1972.

\bibitem{article:Kraichnan:1957}
R.~Kraichnan, Relation of fourth order to second order moments in stationary
  isotropic turbulence, Phys. Rev. 107 (1957) 1485--1490.

\bibitem{article:Kraichnan:1958}
R.~Kraichnan, Irreversible statistical mechanics of incompressible
  hydromagnetic turbulence, Phys. Rev. 109 (1958) 1407--1422.

\bibitem{article:Kraichnan:1959}
R.~Kraichnan, The structure of isotropic turbulence at the very high {R}eynolds
  numbers, J. Fluid. Mech. 5 (1959) 497--543.

\bibitem{article:Kraichnan:1964}
R.~Kraichnan, {K}olmogorov's hypothesis and {E}ulerian turbulence theory, Phys.
  Fluids 7 (1964) 1723--1734.

\bibitem{article:Kraichnan:1965}
R.~Kraichnan, Lagrangian history closure approximation for turbulence, Phys.
  Fluids 8 (1965) 575--598.

\bibitem{article:Kraichnan:1966}
R.~Kraichnan, Isotropic turbulence and inertial range structure, Phys. Fluids 9
  (1966) 1728--1752.

\bibitem{article:Hopf:1952}
E.~Hopf, Statistical hydromechanics and functionals calculus, J. Ratl. Mech.
  Anal. 1 (1952) 87--123.

\bibitem{article:Rosen:1960}
G.~Rosen, Turbulence theory and functional integration. {I}, Phys. Fluids 3
  (1960) 519--524.

\bibitem{article:Rosen:1960:1}
G.~Rosen, Turbulence theory and functional integration. {II}, Phys. Fluids 3
  (1960) 525--528.

\bibitem{article:Novikov:1965}
E.~Novikov, Functionals and the random force method in turbulence, Sov. Phys.
  JETP 20 (1965) 1290.

\bibitem{article:Yanovskii:1976}
S.~Moiseev, A.~Tur, V.~Yanovskii, Spectra and expectation methods of turbulence
  in a compressible fluid, Sov. Phys. JETP 44 (1976) 556--561.

\bibitem{article:Sazontov:1979}
A.~Sazontov, The similarity relation and turbulence spectra in a stratified
  medium, Izv. Atmos. Ocean. Phys. 15 (1979) 566--570.

\bibitem{article:Kraichnan:1962:1}
R.~Lewis, R.~Kraichnan, A space-time functional formalism for turbulence, Comm.
  Pure Appl. Math. 15 (1962) 397--411.

\bibitem{article:Wyld:1961}
H.~Wyld, Formulation of the theory of turbulence in an incompressible fluid,
  Ann. Phys. 14 (1961) 143--165.

\bibitem{article:Rose:1973}
P.~Martin, E.~Siggia, H.~Rose, Statistical dynamics of classical systems, Phys.
  Rev. A 8 (1973) 423--437.

\bibitem{article:Phythian:1977}
R.~Phythian, The functional formalism of classical statistical dynamics, J.
  Phys. A 10 (1977) 777--789.

\bibitem{article:Pavlovic:2002}
N.~Katz, N.~Pavlovic, A cheap {C}affarelli-{K}ohn-{N}irenberg inequality for
  the {N}avier-{S}tokes equation with hyper-dissipation, Geom. Func. Anal. 12
  (2002) 355--379.

\bibitem{article:Li:2005}
Y.~Li, On the true nature of turbulence, [math.AP/0507254] (2005).

\bibitem{lect:Procaccia:1994}
V.~L'vov, I.~Procaccia, Exact resummations in the theory of hydrodynamic
  turbulence: {P}art 0. {L}ine-resummed diagrammatic perturbation approach, in:
  F.~David, P.~Ginsparg (Eds.), Fluctuating Geometries in Statistical Mechanics
  and Field Theory, Proceedings of the {L}es {H}ouches 1994 {S}ummer {S}chool
  of Theoretical Physics, North-Holland, Amsterdam, 1994.

\bibitem{article:Eyink:1996:2}
G.~Eyink, Turbulence noise, J. Stat. Phys. 83 (1996) 955--1019.

\bibitem{article:Andersen:2000}
H.~Andersen, Functional and graphical methods for classical statistical
  dynamics. {I}. {A} formulation of the {M}artin-{S}iggia-{R}ose method, J.
  Math. Phys. 41 (2000) 1979--2020.

\bibitem{article:Yakhot:1981}
V.~Yakhot, Ultraviolet dynamic renormalization group: {S}mall-scale properties
  of a randomly-stirred fluid, Phys. Rev. A 23 (1981) 1486--1497.

\bibitem{article:Lvov:1987}
V.~Belinicher, V.~L'vov, A scale invariant theory of fully developed
  hydrodynamic turbulence, Sov. Phys. JETP 66 (1987) 303--313.

\bibitem{article:Lvov:1991}
V.~L'vov, Scale invariant theory of fully developed hydrodynamic
  turbulence--{H}amiltonian approach, Phys. Rep. 207 (1991) 2--47.

\bibitem{article:Procaccia:1995:2}
V.~L'vov, I.~Procaccia, Exact resummations in the theory of hydrodynamic
  turbulence. {II}. {A} ladder to anomalous scaling, Phys. Rev. E 52 (1995)
  3858--3875.

\bibitem{article:Procaccia:1996}
V.~L'vov, I.~Procaccia, Exact resummations in the theory of hydrodynamic
  turbulence. {III}. {S}cenarios for anomalous scaling and intermittency, Phys.
  Rev. E 53 (1996) 3468--3490.

\bibitem{article:Procaccia:1996:1}
V.~L'vov, I.~Procaccia, Fusion rules in turbulent systems with flux
  equilibrium, Phys. Rev. Lett. 76 (1996) 2898--2901.

\bibitem{article:Procaccia:1996:2}
V.~L'vov, I.~Procaccia, Viscous lengths in hydrodynamic turbulence are
  anomalous scaling functions, Phys. Rev. Lett. 77 (1996) 3541--3544.

\bibitem{article:Procaccia:1996:3}
V.~L'vov, I.~Procaccia, Towards a nonperturbative theory of hydrodynamic
  turbulence: fusion rules, exact bridge relations, and anomalous viscous
  scaling functions, Phys. Rev. E 54 (1996) 6268--6284.

\bibitem{article:Procaccia:1997}
V.~L'vov, E.~Podivilov, I.~Procaccia, Temporal multiscaling in hydrodynamic
  turbulence, Phys. Rev. E 55 (1997) 7030--7035.

\bibitem{article:Procaccia:1998}
V.~L'vov, I.~Procaccia, Computing the scaling exponents in fluid turbulence
  from first principles: {T}he formal setup, Physica A 257 (1998) 165--196.

\bibitem{article:Procaccia:1998:1}
V.~Belinicher, V.~L'vov, I.~Procaccia, A new approach to computing the scaling
  exponents in fluid turbulence from first principles, Physica A 254 (1998)
  215--230.

\bibitem{article:Procaccia:1998:2}
V.~Belinicher, V.~L'vov, A.~Pomyalov, I.~Procaccia, Computing the scaling
  exponents in fluid turbulence from first principles: {D}emonstration of
  multiscaling, J. Stat. Phys. 93 (1998) 797--832.

\bibitem{article:Procaccia:2000}
V.~L'vov, I.~Procaccia, Analytic calculation of the anomalous exponents in
  turbulence: using the fusion rules to flush out a small parameter, Phys. Rev.
  E 62 (2000) 8037--8057.

\bibitem{article:Procaccia:1999}
I.~Arad, V.~L'vov, I.~Procaccia, Correlation functions in isotropic and
  anisotropic turbulence: {T}he role of the symmetry group, Phys. Rev. E 59
  (1999) 6753--6765.

\bibitem{article:Procaccia:2005}
L.~Biferale, I.~Procaccia, Anisotropy in turbulent flows and in turbulent
  transport, Phys. Rep. 414 (2005) 43--164.

\bibitem{book:McComb:1990}
W.~McComb, The physics of fluid turbulence, Clarendon Press, Oxford, 1990.

\bibitem{article:Woodruff:1998}
L.~Smith, S.~Woodruff, Renormalixation-group analysis of turbulence, Ann. Rev.
  Fluid Mech. 30 (1998) 275--310.

\bibitem{article:Kraichnan:1987}
R.~Kraichnan, An interpretation of the yakhot-orzag turbulence, Phys. Fluids 30
  (1987) 2400--2405.

\bibitem{article:Eyink:1994}
G.~Eyink, The renormalization group method in statistical hydrodynamics, Phys.
  Fluids 6 (1994) 3063--3078.

\bibitem{article:Eyink:1993}
G.~Eyink, Lagrangian field theory, multifractals, and universal scaling in
  turbulence, Phys. Lett. A 172 (1993) 335--360.

\bibitem{article:Eyink:1993:1}
G.~Eyink, Renormalization group and operator product expansion in turbulence:
  Shell models, Phys. Rev. E 48 (1993) 1823--1838.

\bibitem{article:Giles:2001}
M.~Giles, Anomalous scaling in homogeneous isotropic turbulence, J. Phys. A 34
  (2001) 4389--4435.

\bibitem{book:Temam:1999}
T.~Dubois, F.~Jauberteau, R.~Temam, Dynamic multilevel methods and the
  numerical simulation of turbulence, Cambridge University Press, Cambridge,
  1999.

\bibitem{lect:Galdi:2000}
G.~Galdi, An introduction to the {N}avier-{S}tokes initial-boundary value
  problem, in: G.~Galdi, J.~Heywood, R.~Rannacher (Eds.), Fundamental
  Directions in Mathematical Fluid Mechanics, Advances in Mathematical Fluid
  Mechanics, Birkh\"auser Verlag, Basel, 2000.

\bibitem{article:Pavlik:1990}
M.~Altaisky, S.~Moiseev, S.~Pavlik, Scaling and supersymmetry in spectral
  problems of strong turbulence, Phys. Lett. A 147 (1990) 142--146.

\bibitem{article:Gozzi:1988}
E.~Gozzi, Hidden brs invariance in classical mechanics, Phys. Lett. B 201
  (1988) 525--528.

\bibitem{article:Thacker:1989}
E.~Gozzi, M.~Reuter, W.~Thacker, Hidden brs invariance in classical mechanics.
  {II}, Phys. Rev. D 40 (1989) 3363--3377.

\bibitem{article:Reuter:1994}
E.~Gozzi, M.~Reuter, Lyapunov exponents, path integrals and forms, Chaos,
  Solitons and Fractals 4 (1994) 1117--1139.

\bibitem{article:Thacker:1997}
W.~Thacker, A path integral for turbulence in incompressible fluids, J. Math.
  Phys. 38 (1997) 300--320.

\bibitem{article:Frisch:1991}
U.~Frisch, From global scaling, a la {K}olmogorov, to local multifractal
  scaling in fully developed turbulence, Proc. R. Soc. Lond. A 434 (1991)
  89--99.

\bibitem{book:Yaglom:1975}
A.~Monin, A.~Yaglom, Statistical Fluid Mechanics, the MIT Press, Cambridge MA,
  1975.

\bibitem{article:Uno:2003}
Y.~Kaneda, T.~Ishihara, M.~Yokokawa, K.~Itakura, A.~Uno, Energy dissipation
  rate and energy spectrum in high resolution direct numerical simulations of
  turbulence in a periodic box, Phys. Fluids 15 (2003) L21--L24.

\bibitem{article:Hill:2002:1}
R.~Hill, The approach of turbulence to the locally homogeneous asymptote as
  studied using exact structure-function equations, [physics/0206034] (2002).

\bibitem{lfthesis:2006}
E.~Gkioulekas, A theoretical study of the cascades of {3D}, {2D}, and {QG}
  turbulence, Ph.D. thesis, University of Washington, (Advisor: Ka-Kit Tung)
  (2006).

\bibitem{article:Rasmussen:1999}
H.~Rasmussen, A new proof of {K}olmogorov's 4/5-law, Phys. Fluids 11 (1999)
  3495--3498.

\bibitem{article:Monin:1959}
A.~Monin, The theory of locally isotropic turbulence, Dokl. Akad. Nauk. SSSR
  125 (1959) 515--518.

\bibitem{article:Lindborg:1996}
E.~Lindborg, A note on {K}olmogorov's third order structure function law, the
  local isotropy hypothesis and the pressure velocity correlation, J. Fluid.
  Mech. 326 (1996) 343--356.

\bibitem{article:Hill:1997}
R.~Hill, Applicability of {K}olmogorov's and {M}onin's equations of turbulence,
  J. Fluid. Mech. 353 (1997) 67--81.

\bibitem{submitted:Tung:2003}
E.~Gkioulekas, K.~Tung, On the double cascades of energy and enstrophy in two
  dimensional turbulence. {P}art 1. {T}heoretical formulation, Discrete and
  Continuous Dynamical Systems B 5 (2005) 79--102.

\bibitem{submitted:Tung:2003:1}
E.~Gkioulekas, K.~Tung, On the double cascades of energy and enstrophy in two
  dimensional turbulence. {P}art 2. {A}pproach to the {KLB} limit and
  interpretation of experimental evidence, Discrete and Continuous Dynamical
  Systems B 5 (2005) 103--124.

\bibitem{article:Hill:2002}
R.~Hill, Exact second-order structure function relationships, J. Fluid. Mech.
  468 (2002) 317--326.

\bibitem{article:Robert:2000}
J.~Duchon, R.~Robert, Inertial range dissipation for weak solitions of
  incompressible euler and navier-stokes equations, Nonlinearity 13 (2000)
  249--255.

\bibitem{article:Eyink:2003:1}
G.~Eyink, Local $4/5$-law and energy dissipation anomaly in turbulence,
  Nonlinearity 16 (2003) 137--145.

\bibitem{article:Eyink:2003:2}
M.~A. Taylor, S.~K. amd G.L.~Eyink, Recovering isotropic statistics in
  turbulence simulations: The kolmogorov $4/5$th law, Phys. Rev. E 68 (2003)
  026310.

\bibitem{article:Tanveer:1999}
Q.~Nie, S.~Tanveer, A note on the third-order structure functioms in
  turbulence, Proc. R. Soc. Lond. A 455 (1999) 1615--1635.

\bibitem{article:Lindborg:1999:1}
E.~Lindborg, Correction to the four-fifths law due to variations of the
  dissipation, Phys. Fluids 11 (1999) 510--512.

\bibitem{article:Antonia:2001}
L.~Danaila, F.~Anselmet, T.~Zhou, R.~A. Antonia, Turbulent energy scale budget
  equations in a fully developed channel flow, J. Fluid. Mech. 430 (2001)
  87--109.

\bibitem{article:Antonia:2002}
L.~Danaila, F.~Anselmet, R.~A. Antonia, An overview of the effect of
  large-scale inhomogeneities on small-scale turbulence, Phys. Fluids 14 (2002)
  2475--2484.

\bibitem{article:Zhou:2004}
L.~Danaila, F.~Anselmet, T.~Zhou, Turbulent energy scale-budget equations for
  nearly homogeneous sheared turbulence, Flow, Turbulence, and Combustion 72
  (2004) 287--310.

\bibitem{article:Burattini:2004}
L.~Danaila, R.~A. Antonia, P.~Burattini, Progress in studying small-scale
  turbulence using exact two-point equations, New J. Phys 6 (2004) 128.

\bibitem{article:Lebedev:1994:2}
V.~L'vov, V.~Lebedev, Anomalous scaling and fusion rules in hydrodynamic
  turbulence, [chao-dyn/940003] (1994).

\bibitem{book:Hibbs:1965}
R.~Feynman, A.~Hibbs, Quantum mechanics and path integrals, McGraw-Hill Book
  Co., New York, 1965.

\bibitem{book:Stevens:1995}
C.~Stevens, The six core theories of modern physics, the MIT Press, Cambridge
  Mass., 1995.

\bibitem{article:Vergassola:2000}
G.~Boffetta, A.~Celani, M.~Vergassola, Inverse energy cascade in two
  dimensional turbulence: deviations from gaussian behavior, Phys. Rev. E 61
  (2000) 29--32.

\bibitem{article:Borue:1994}
V.~Borue, Inverse energy cascade in stationary two dimensional homogeneous
  turbulence, Phys. Rev. Lett. 72 (1994) 1475--1478.

\bibitem{article:Gurarie:2001}
S.~Danilov, D.~Gurarie, Non-universal features of forced two dimensional
  turbulence in the energy range, Phys. Rev. E 63 (2001) 020203.

\bibitem{article:Gurarie:2001:1}
S.~Danilov, D.~Gurarie, Forced two-dimensional turbulence in spectral and
  physical space, Phys. Rev. E 63 (2001) 061208.

\bibitem{article:Danilov:2003}
S.~Danilov, Non-universal features of forced 2d turbulence in the energy and
  enstrophy ranges, Discrete and Continuous Dynamical Systems B 5 (2005)
  67--78.

\bibitem{article:Falkovich:1994:1}
G.~Falkovich, Bottleneck phenomenon in developed turbulence, Phys. Fluids 6
  (1994) 1411--1414.

\bibitem{article:Tung:2005}
E.~Gkioulekas, K.~Tung, On the double cascades of energy and enstrophy in two
  dimensional turbulence. {P}art 1. {T}heoretical formulation, Discrete and
  Continuous Dynamical Systems B 5 (2005) 79--102.

\bibitem{article:Tung:2005:1}
E.~Gkioulekas, K.~Tung, On the double cascades of energy and enstrophy in two
  dimensional turbulence. {P}art 2. {A}pproach to the {KLB} limit and
  interpretation of experimental evidence, Discrete and Continuous Dynamical
  Systems B 5 (2005) 103--124.

\bibitem{article:Fischer:2005}
P.~Fischer, Multiresolution analysis for 2d turbulence. part 1: Wavelets vs
  cosine packets, a comparative study, Discrete and Continuous Dynamical
  Systems B 5 (2005) 659--686.

\bibitem{article:Weichman:1995}
C.~Mou, P.~Weichman, Multicomponent turbulence, the spherical limit, and
  non-kolmogorov spectra, Phys. Rev. E 52 (1995) 3738--3796.

\bibitem{article:Donsker:1964}
M.~Donsker, On function space integrals, in: W.~Martin, I.~Segal (Eds.),
  Analysis in Function Space, MIT Press, Cambridge, MA, 1964, pp. 17--30.

\bibitem{article:Furutsu:1963}
K.~Furutsu, On the statistical theory of electromagnetic waves in a fluctuating
  medium, J. Res. Nat. Bur. Standards D 67 (1963) 303--323.

\end{thebibliography}
\bibliographystyle{elsart-num}

\end{document}